\documentclass[fleqn,usenatbib]{mnras}

\usepackage{newtxtext,newtxmath}

\usepackage[T1]{fontenc}

\DeclareRobustCommand{\VAN}[3]{#2}
\let\VANthebibliography\thebibliography
\def\thebibliography{\DeclareRobustCommand{\VAN}[3]{##3}\VANthebibliography}

\usepackage{graphicx}	
\usepackage{amsmath}	
\usepackage{caption}
\usepackage{color}
\usepackage{hyperref}
\usepackage{lipsum}
\usepackage{ulem}
\usepackage{soul}

\newcommand{\lognh}[1]{{$\log N_{\rm H}\rm\ [cm^{-2}]#1$}}	
\newcommand{\loglx}[1]{{$\log L_{\rm X}\rm\ [erg\ s^{-1}]#1$}}	
\newcommand{\um}{$\rm \mu m$ }

\title[Observational Properties of AGN Obscuration]{Observational Properties of AGN Obscuration During the Peak of Accretion Growth}

\author[Bovornpratch Vijarnwannaluk]{
Bovornpratch Vijarnwannaluk,$^{1,6}$\thanks{E-mail: bovornpratch.v@astr.tohoku.ac.jp}
Masayuki Akiyama,$^{1}$
Malte Schramm,$^{2}$
Yoshihiro Ueda,$^{3}$
\newauthor
Yoshiki Matsuoka,$^{4}$
Yoshiki Toba,$^{4,5,6}$
Naoki Matsumoto,$^{1}$
Angel Ruiz,$^{7}$
Ioannis Georgantopoulos,$^{7}$
\newauthor
Ektoras Pouliasis,$^{7}$
Elias Koulouridis ,$^{7}$
Kohei Ichikawa$^{1,8}$
Marcin Sawicki$^{9}$
Stephen Gwyn$^{10}$
\\
$^{1}$Astronomical Institute, Tohoku University, Aramaki, Aoba-ku, Sendai, Miyagi 980-8578, Japan\\
$^{2}$Graduate School of Science and Engineering, Saitama University, 255 Shimo-Okubo, Sakura-ku, Saitama City, Saitama 338-8570, Japan\\
$^{3}$Department of Astronomy, Kyoto University, Kitashirakawa-Oiwake-cho, Sakyo-ku, Kyoto 606-8502, Japan \\
$^{4}$Research Center for Space and Cosmic Evolution, Ehime University, 2-5 Bunkyo-cho, Matsuyama, Ehime 790-8577, Japan \\
$^{5}$National Astronomical Observatory of Japan, 2-21-1 Osawa, Mitaka, Tokyo 181-8588, Japan \\
$^{6}$Academia Sinica Institute of Astronomy and Astrophysics, 11F of Astronomy-Mathematics Building, \\ AS/NTU, No.1, Section 4, 12 Roosevelt Road, Taipei 10617, Taiwan \\
$^{7}$Institute for Astronomy Astrophysics Space Applications and Remote Sensing (IAASARS), National Observatory of Athens, \\ I. Metaxa \& V. Pavlou, Penteli 15236, Greece \\
$^{8}$Faculty of Science and Engineering, Waseda University, 3-4-1 Okubo,
Shinjuku, Tokyo 169-8555, Japan \\
$^{9}$ Institute for Computational Astrophysics and Department of Astronomy and Physics, Saint Mary’s University, Halifax, NS B3H 3C3, Canada \\
$^{10}$ NRC-Herzberg, 5071 West Saanich Road, Victoria, British Columbia V9E 2E7, Canada
}

\date{Accepted XXX. Received YYY; in original form ZZZ}

\pubyear{2024}

\begin{document}
\label{firstpage}
\pagerange{\pageref{firstpage}--\pageref{lastpage}}
\maketitle

\begin{abstract}
We investigated the gas obscuration and host galaxy properties of active galactic nuclei (AGN) during the peak of cosmic accretion growth of supermassive black holes (SMBHs) at redshift 0.8-1.8 using X-ray detected AGN with mid-infrared and far-infrared detection. The sample was classified as type-1 and type-2 AGN using optical spectral and morphological classification while the host galaxy properties were estimated with multiwavelength SED fitting. For type-1 AGN, the black hole mass was determined from MgII emission lines while the black hole mass of type-2 AGN was inferred from the host galaxy's stellar mass.  Based on the derived parameters, the distribution of the sample in the absorption hydrogen column density ($N_{\rm H}$) vs. Eddington ratio diagram is examined. Among the type-2 AGN, $28\pm5$\% are in the forbidden zone, where the obscuration by dust torus cannot be maintained due to radiation pressure on dusty material. The fraction is higher than that observed in the local universe from the BAT AGN Spectroscopic Survey (BASS) data release 2  ($11\pm3$\%). The higher fraction implies that the obscuration of the majority of AGN is consistent with the radiation pressure regulated unified model but with an increased incidence of interstellar matter (ISM) obscured AGN. We discuss the possibility of dust-free absorption in type-1 AGN and heavy ISM absorption in type-2 AGN. We also find no statistical difference in the star-formation activity between type-1 and type-2 AGN which may suggest that obscuration triggered by a gas-rich merging is not common among X-ray detected AGN in this epoch.
\end{abstract}




\section{Introduction}
Active galactic nuclei (AGN) can be broadly classified into type-1 and type-2 AGN where broad emission lines and a blue power-law continuum present among type-1 AGN  but disappear among type-2 AGN. These observations led to the development of the unified model of AGN \citep{1993ARA&A..31..473A,1995PASP..107..803U} which explains the spectral difference as a result of different lines of sight towards the AGN with a portion of the lines-of-sight obscured by a dusty torus.

Recent studies of AGN obscuration suggest that the dusty torus is regulated by radiation pressure from the AGN \citep{2006MNRAS.373L..16F,2008MNRAS.385L..43F,2009MNRAS.394L..89F,2012ApJ...758...66W,2016ApJ...828L..19W}.  Due to the large cross-section of dust particles to UV radiation, the critical luminosity which balances with the gravitational forces of the black hole is lower than the typical Eddington luminosity ($L_{\rm Edd}$). The effective Eddington ratio, which describes the critical Eddington ratio of dusty gas separates the $N_{\rm H}-\lambda_{\rm Edd}$ distribution into 2 regions.  In the first region, the line-of-sight dusty obscuration is stable hence obscuration is long-lived. The second region is the forbidden zone where the radiation pressure on dust particles overcomes the gravitational force of the black hole and drives dusty outflows which clear the line of sight, changing obscured AGN to unobscured \citep{2021ApJ...906...21J,2022ApJ...938...67R}. In addition to the two regions, there is a parameter space where the X-ray absorption is due to interstellar matter (ISM) \citep{1995ApJ...454...95M}, with a fiducial hydrogen column density of \lognh{\sim22}.

This model has been successful in describing the properties of local hard X-ray selected AGN, which shows that obscured AGN tend to avoid the forbidden zone \citep{2009MNRAS.394L..89F,2017Natur.549..488R}. In contrast, AGN which contains dusty outflows tends to reside in the forbidden zone \citep{2016A&A...592A.148K,2021ApJS..257...61Y,2022MNRAS.517.3377S}.  Among Compton-thin AGN (CTN-AGN; \lognh{=20-24}), the fraction of obscured AGN (\lognh{=22-24}) shows a decreasing trend with increasing AGN luminosity \citep{2003ApJ...598..886U,2011ApJ...728...58B}. Hydrodynamic simulations which apply the concept of radiation pressure regulation (eg. \citet{2012ApJ...758...66W,2015ApJ...812...82W,2016ApJ...828L..19W}) could reproduce structures of accretion flows analogous to a torus structure and qualitatively reproduce the trends in the obscured fraction of AGN with luminosity. 

On the other hand, our understanding of obscuration properties beyond the local universe is limited. Compton-thin AGN shows a second trend where the fraction of obscured AGN increases towards high redshift \citep{2005ApJ...635..864L,2008A&A...490..905H,2014ApJ...786..104U,2015MNRAS.451.1892A,2015ApJ...802...89B,2022ApJ...941...97V,2023ApJ...943..162P,2023A&A...676A..49S}.  Recently, this trend is also observed in samples of multi-wavelength selected AGN \citep{2022ApJ...941..191L,2023arXiv231012330L}. Above cosmic noon, some studies suggest that the trend of the obscured fraction with AGN luminosity flattens, indicating that many luminous AGN can be obscured \citep{2014MNRAS.445.3557V,2018MNRAS.473.2378V,2022ApJ...941...97V,2023ApJ...943..162P,2024arXiv240113515P}.

The increasing trend with redshift suggests that an additional component of obscuration beyond the dusty torus is required. One possibility is an increase incidence of ISM obscuration \citep{1995ApJ...454...95M,2017MNRAS.464.4545B,2017MNRAS.465.4348B,2022A&A...666A..17G, 2023MNRAS.tmp.3140A,2024MNRAS.527L.144A}. \citet{2022A&A...666A..17G} showed that due to the cosmological evolution of the galaxy ISM and size of galaxies \citep{2010ApJ...713..686D,2013ARA&A..51..105C,2014ApJ...788...28V,2019ApJ...880...57M,2018ApJ...853..179T}, incidence of ISM obscuration can occur with larger column densities (\lognh{>22}) than the typical amount observed in the local universe. 

An alternative scenario of obscuration may be due to abundant gas-rich mergers beyond the local universe. Hydro-dynamical simulations of galaxy mergers suggested that massive inflows of gas and dust are funnelled towards the nuclear regions due to the gravitational disturbance \citep{2006ApJS..163....1H,2008ApJS..175..356H,2018MNRAS.478.3056B}. During this phase, AGN activity occurs under heavy obscuration before strong AGN outflows clear the line of sight toward the observer and the AGN becomes unobscured. Often referred to as the evolutionary scenario of major mergers,
it suggests fundamental differences between unobscured and obscured AGN which may be reflected in the AGN or host galaxy properties, since the obscured phase precedes the unobscured AGN phase \citep{2011ApJ...731..117H,2006ApJ...639...37K,2013A&A...552A.135K,2014A&A...570A..72K,2018ARA&A..56..625H}.  In the local universe, such a scenario of obscuration is seen in ultraluminous infrared galaxies (ULIRGs) \citep{1996ARA&A..34..749S}. Defined as galaxies with infrared luminosity exceeding $L_{\rm IR}>10^{12}\ L_{\odot}$, they represent the luminous end of the infrared luminosity function. Previous studies show that most ULIRGs contain heavily obscured luminous AGN and host galaxies are in a late-stage gas-rich merger with high star-formation \citep{2008PASJ...60S.489I,2011MNRAS.410..573G,2017ApJ...835...36T,2014ApJ...794..139I,2017MNRAS.468.1273R,2021MNRAS.506.5935R,2021ApJS..257...61Y,2022ApJ...936..118Y}. Similar results were also reported in \citet{2006ApJ...651...93K} for bright {\it IRAS} galaxies (BIRGs) with infrared luminosities $L_{\rm IR}$ between $10^{10}$ and $10^{12}\ L_{\odot}$. Therefore, the increasing fraction of obscured AGN at high redshift may be a result of an increased frequency of AGN triggered by galaxy mergers.

At cosmic noon ($z\sim1-3$), the discussion of the obscuration properties (eg. $N_{\rm H}-\lambda_{\rm Edd}$ diagram) is focused on specific samples of AGN, such as highly luminous reddened quasars \citep{2021ApJ...906...21J,2020MNRAS.495.2652L} or X-ray obscured broad-line AGN \citep{2018MNRAS.479.5022L}. Whereas, the discussion for the typical population of AGN near the knee of the luminosity function (\loglx{=43.5-44.5}) is limited. AGN near the knee of the luminosity function contributes to the bulk of the black hole mass density \citep{2014ApJ...786..104U}. Thus, understanding this population may shed some light on the relation between AGN obscuration, the host galaxy, and cosmic black hole growth.

In this study, we investigated the origin of AGN obscuration at redshift 0.8-1.8 during the peak accretion history of AGNs near the knee of the X-ray luminosity function (\loglx{=43.5-45}) by comparing with the obscuration properties in the local universe seen in the $N_{\rm H}-\lambda_{\rm Edd}$ diagram. In \citet{2022ApJ...941...97V} (hereafter referred to as V22), we showed that in addition to a large obscured fraction, the optical/IR SED of unobscured and obscured AGN show a large scatter in the SED shapes, which may indicate multiple processes of obscuration. Here, we used a sample of type-1 and type-2 AGN drawn from the deep and wide multiwavelength dataset in the Hyper Suprime Cam (HSC) deep survey of the XMM-LSS field \citep{2004JCAP...09..011P}. We used the rich multiwavelength photometric and spectroscopic datasets to estimate the AGN and host galaxy properties by performing optical spectroscopic and SED fitting analyses.  By assuming a virial mass relation and a $M_{\rm stellar}-M_{\rm BH}$ relationship, we estimated the black hole mass for type-1 and type-2 AGN, respectively. We investigate the star formation rate of type-1 and type-2 AGN which may be indicative of ISM obscuration induced by a gas-rich merger as opposed to ISM obscuration due to the cosmological evolution of galaxies. Here, we adopt the metal abundances of \citet{1989GeCoA..53..197A}, the galactic hydrogen column density in the area  of $N_{\rm H}=3.57\times10^{20} \rm  cm^{-2}$ \citep{2018MNRAS.478.2132C} and a flat standard $\rm \Lambda CDM$ cosmology with $H_{\rm 0} = 70\ \rm km\ s^{-1}\ Mpc^{-1}$,$\Omega_{\rm M} = 0.3$, and $\Omega_{\rm \lambda} = 0.7$ was assumed. Magnitudes are reported in the AB magnitude system.

\section{Data} \label{sec:data}
\subsection{X-ray AGN Catalogue} \label{sec:xagncat}
The parent sample of X-ray AGN was taken from V22,  which contains 3462 X-ray selected AGNs within the HSC-Deep XMM-LSS survey area. These AGN were detected with the \textit{XMM-Newton} telescope using the PN, MOS1, and MOS2 detectors in 0.5-2, 2-10, and 0.5-10 keV bands as part of the XMM-SERVS point source catalogue \citep{2018MNRAS.478.2132C} in the XMM-LSS. The X-ray survey depth in 0.5-2, 2-10, and 0.5-10 keV is $1.7\times10^{-15}$, $1.3\times10^{-14}$, and $6.5\times10^{-15}\ \rm erg\ cm^{-2}\ s^{-1}$, respectively. 

V22 presented point spread function (PSF) convolved photometry measured from deep optical and infrared photometric surveys in the HSC-Deep XMM-LSS survey field for a survey area of 3.5 $\rm Deg^2$. This includes deep HSC imaging in $grizy$ bands, the Canada France Hawaii Telescope large area $U$-band deep survey (CLAUDS; \citealt{2019MNRAS.489.5202S}), VISTA deep extragalactic observations survey (VIDEO; \citealt{2013MNRAS.428.1281J}), and Spitzer extragalactic representative volume survey (SERVS, \citealt{2012PASP..124..714M}). For the HSC datasets, the image data from HSC were reduced by the HSC collaboration using the HSC pipeline \citep{2017ASPC..512..279J,2018PASJ...70S...5B,2019ASPC..523..521B,2019ApJ...873..111I} while the photometry and astrometry were calibrated against the first data release of the Panoramic survey telescope and rapid response system (Pan-STARRS1; \citealt{2012ApJ...756..158S,2012ApJ...750...99T,2013ApJS..205...20M,2016arXiv161205560C,2020ApJS..251....6M}). The survey depth reaches $i$-mag $\sim$ 26.9, J-mag $\sim$ 24.44, and 3.6 \um depth of 2 $\rm \mu Jy$.  We refer the readers to Table 1 of V22 for the reported survey depth in each band and the details on the construction of the PSF-convolved photometry. With the PSF-convolved photometry, we can obtain accurate AGN colours, which is crucial for photometric redshift estimation of AGN that does not have spectroscopic information. V22 calculated the photometric redshift using  LePHARE \citep{1999MNRAS.310..540A,2006A&A...457..841I} with galaxy, AGN, and composite AGN models from \citet{2009ApJ...690.1250S}. The final redshift completeness among the 3542 AGN is 97\%(3458) and with 37\%(1321) spectroscopic redshifts.

\subsection{Additional Far-infrared} \label{sec:sdss_spec}

Additional mid-infrared and far-infrared datasets from the Spitzer data fusion catalogue \citep{2015fers.confE..27V} were added to decompose the AGN hot dust emission and estimate the star-formation rate. The Spitzer data fusion catalogue contains IRAC 3.6-8.0 \um and MIPS 24$\rm \mu m$ aperture photometry. It also contains Herschel PACS \citep{2010A&A...518L...2P} and SPIRE \citep{2010A&A...518L...3G} 250, 350, and 500 $\mu m$ PRF-photometry based on 24 \um prior positions.  Both MIPS24\um and the FIR photometry fully cover the multiwavelength survey area presented in V22. Firstly, the two catalogues were matched using the band merged IRAC channel 1 (3.6 \um) \& 2 (4.5 \um) coordinates and optical counterpart coordinates of V22 with a 2" radius aperture. Among the original 3462 AGN, 1359 AGN have mid-infrared counterparts with 24 $\mu m$ photometry. For the remaining objects, no 24 \um detection (and subsequently, no FIR data) was found in a 5.25" arcsecond aperture.  Visual inspection of the FIR images was performed to examine the quality of the far-infrared datasets. After adding the confusion noise to the photometric noise in the quadrature, we determined that sources with S/N less than three show no clear detection in the far infrared bands. Therefore, we consider FIR bands with S/N$<3$ as non-detection and replaced the flux with $3\sigma$ upper limits. For 250, 350, and 500 \um bands, the average $3\sigma$ upper limits are 17.0, 13.5, and 13.6 $\rm \mu Jy$, respectively.

\section{Analysis}

\subsection{Sample Selection} \label{sec:analy:samp}
To construct the AGN sample, we first chose 1521 AGN at redshift 0.8-1.8 from the V22 catalogue, then limited the AGN to those that have successfully measured photometry in the i-band using a point spread function (PSF) model, fitted photometry using a cmodel\footnote{The cmodel is a linear combination of an exponential profile and de Vaucouleurs profile convolved with the PSF of HSC \citep{2018PASJ...70S...5B}.} as well as successfully measured 2nd-moment shape measurements in the HSC S21A internal catalogue. Among them, we limit the sample to 676 AGN that have 24\um detection. We limited the redshift range to 0.8-1.8 to ensure the detection of the MgII emission line which is a reliable blackhole mass estimator by optical spectrographs such as BOSS\citep{2013AJ....146...32S} and FOCAS \citep{2002PASJ...54..819K}. In addition, we limited the sample to AGN detected in the 0.5-10 keV band and are not within the same region as local galaxies in the HyperLEDA catalogue \citep{2014A&A...570A..13M}.  We adopted a final sample of 666 AGN for later discussion. Among them, 395 AGN were detected in both 0.5-2 and 2-10 keV bands. 171, 72, and 28 AGN have single band detection in 0.5-2 keV band, 2-10 keV band, and 0.5-10 keV respectively. The spectroscopic redshift completeness is 54\%(363 AGN). We obtained optical spectra from the SDSS DR16 archive \citep{2020ApJS..249....3A} for 239 AGN. At redshift 0.8-1.8, the MgII emission line is expected to fall into the detector with sufficient continuum emission on both sides of the line. Within the sample of 666 AGN, 166 have FIR detection in either one of the SPIRE 250, 350, or 500\um bands. The catalogue listing the properties of the 666 AGN is provided with the supplement material online and the description can be found in appendix \ref{sec:appendix}.

\subsection{Type-1 \& Type-2 Classification}

Classification of AGN depends on the wavelength in discussion. From hereafter,  we refer to type-1 AGN as those with broad MgII, CIII], or CIV emission lines (FWHM$>1000\ \rm km\ s^{-1}$) or with morphology consistent with a point source. Whereas type-2 AGN are those that have extended optical morphology and have no confirmation of broad-emission lines. The AGN is classified as obscured if the Hydrogen column density is between \lognh{=22-24} while unobscured AGN is that with \lognh {<22}. 

To classify the AGN as type-1 or type-2 AGN, we first adopted the spectroscopic classification presented in \citet{2018MNRAS.478.2132C} which adopted the spectroscopic classification from public spectroscopic datasets in the XMM-SERVS area.  In addition to the spectroscopic classification, we also adopted a morphological classification where AGN with point source morphology is considered type-1 AGN.  To classify sources as point sources or extended sources, we used the ratio between the cmodel and PSF flux in HSC $i$-band. Sources whose morphology is close to a point source will have a flux ratio near unity. 

\begin{figure}
    \centering
    \includegraphics[width=\columnwidth]{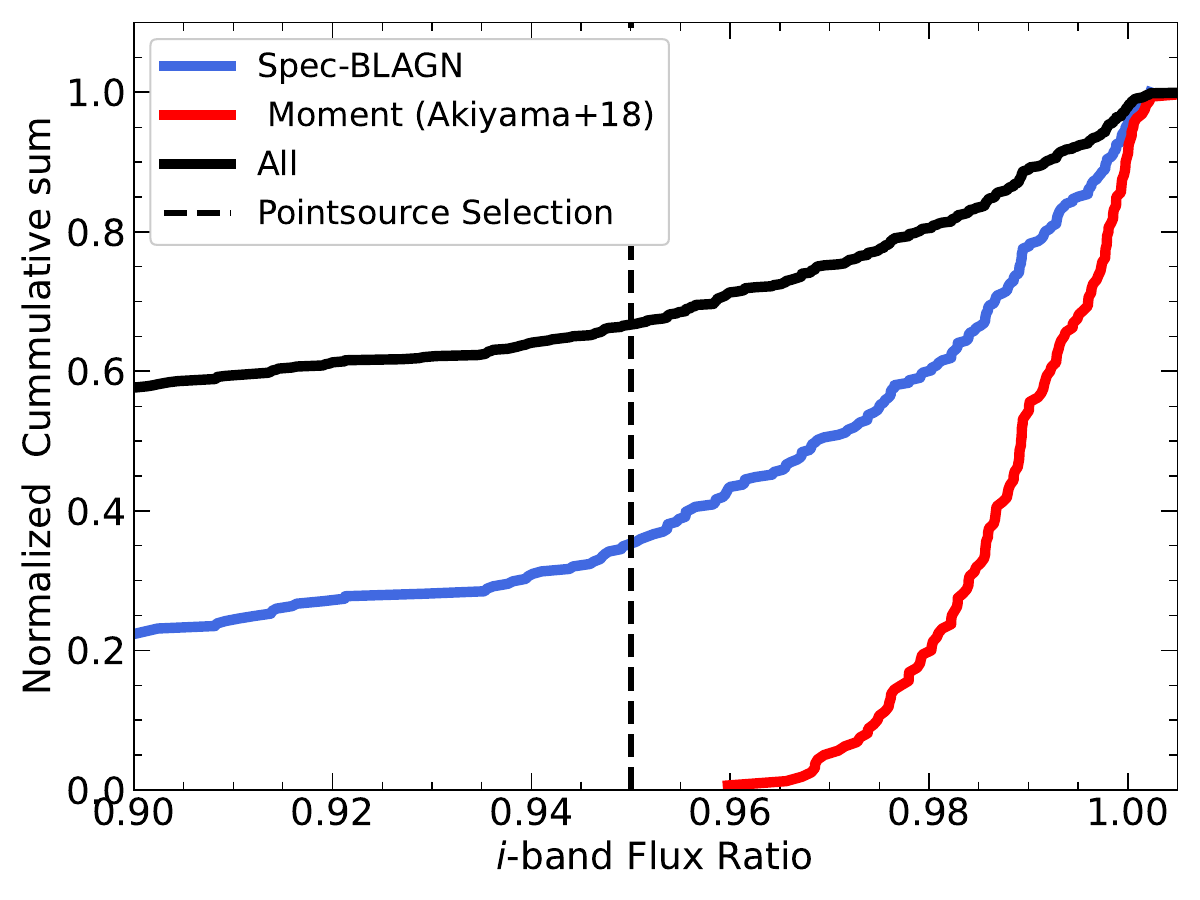}
    \caption{The normalised cumulative distribution of the ratio between $i$-band cmodel flux over PSF flux of AGN at redshift 0.8-1.8. The black line represents all MIPS 24 $\rm\mu m$ detected AGN. The blue line represents spectroscopically confirmed broad-line AGN. The red line represents X-ray detected point sources classified using the 2nd-moment classification presented by \citet{2018PASJ...70S..34A}. The black dashed line shows the flux ratio used to classify point sources in this work.}
    \label{fig:flux_ratio}
\end{figure}

An alternative method to classify point sources was presented in \citet{2018PASJ...70S..34A} using the 2nd moment of the PSF. Figure \ref{fig:flux_ratio} shows the normalised cumulative distribution of the $i$-band flux ratio of AGN at redshift 0.8-1.8, the sample of spectroscopic broad-line AGN, and AGN which were classified as point-sources based on the 2nd moment of the image, respectively. \citet{2018PASJ...70S..34A} showed that the 2nd-moment classification is efficient up to $i=24$ but may miss slightly extended sources due to host galaxy contamination as shown in figure \ref{fig:flux_ratio}. We adopted the i-band flux ratio to recover the slightly extended sources and classify AGN as a point source when the $i$-band flux ratio is larger than 0.95. 36 more AGNs were classified as type-1 AGNs using this method.  The final classification in the sample is 328 type-1 AGN and 338 type-2 AGN. For the type-1 AGN and type-2 AGN samples, 291 and 72 have spectroscopic redshifts from the literature, respectively. The number of spectroscopic redshifts confirmed AGN and FIR detected AGN is summarised in table \ref{tab:t1t2_spec_firdet}.  Among the broad-line AGN, 64\% have point source-like morphology. The remaining 36\% show extended profiles since they are at a lower redshift than the point source broad line AGN.

\begin{table}
    \centering
    \begin{tabular}{l|c|c|c|c|c}
         \hline
         Sample & & \multicolumn{2}{c}{Redshift} & \multicolumn{2}{c}{FIR Detection} \\
         \hline
         {    } & Total & Spec-z & Photo-z & Detected & Non-Detected \\
         \hline
         Type-1 & 328 & 291 & 37 & 75 & 253 \\
         Type-2 & 338 & 72 & 266 & 91 & 247 \\
         \hline
         Total & 666 & 363 & 303 & 166 & 500 \\
         \hline
    \end{tabular}
    \caption{Number of spectroscopic redshifts confirmed AGN and FIR detected AGN among the type-1 and type-2 AGN samples.}
    \label{tab:t1t2_spec_firdet}
\end{table}

\subsection{Black hole mass estimates for broad-line AGNs} \label{sec:analy:bhmass}

The black hole mass ($M_{\rm BH}$) of broad-line AGN can be estimated using single-epoch virial black hole mass relationships \citep{2005ApJ...630..122G,2006ApJ...641..689V,2009ApJ...699..800V} using the  FWHM of the broad-lines and the AGN continuum luminosity. We used the spectral fitting code PyQSOFit \citep{2018ascl.soft09008G} to fit the MgII emission line and the AGN continuum of the SDSS spectra. The AGN continuum was fitted with a power-law model combined with FeII emission lines. We limited the maximum FWHM of the FeII emission lines to 10,000 $\rm km\ s^{-1}$. Wavelengths in the spectra affected by strong optical sky emission lines and telluric absorption were masked.  The MgII emission line was first fitted with a single broad Gaussian component. Additional components were added after a visual examination of the initial fit.

\begin{figure*}
    \centering
    \includegraphics[width=\linewidth]{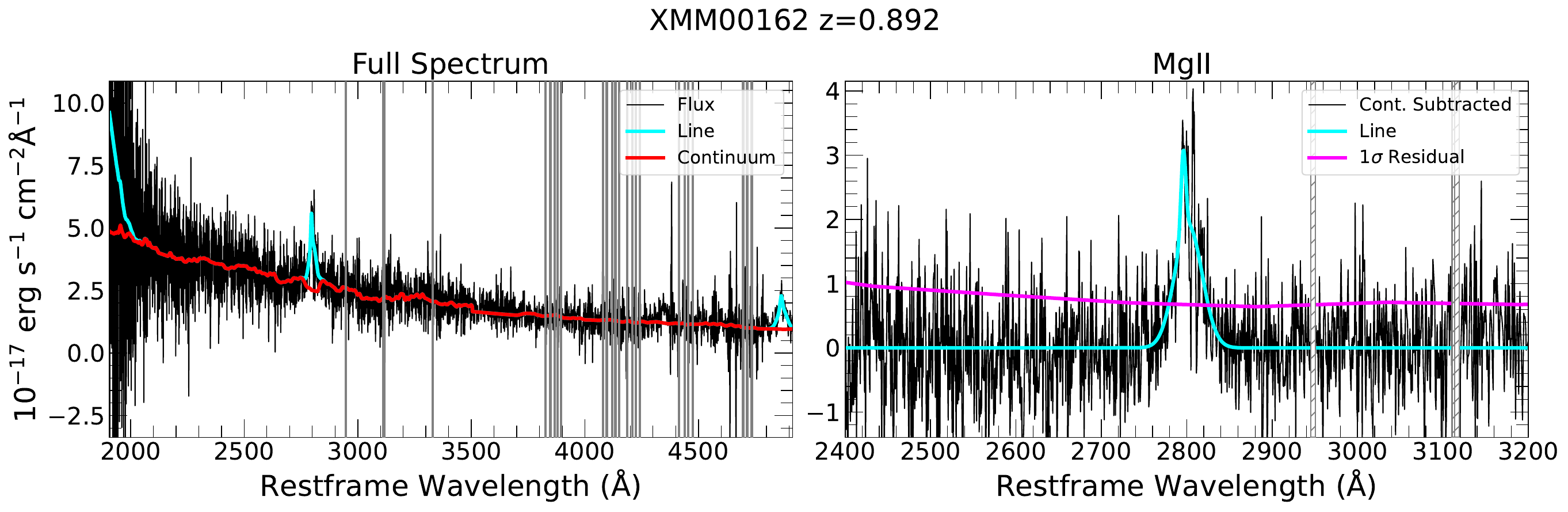}
    \includegraphics[width=\linewidth]{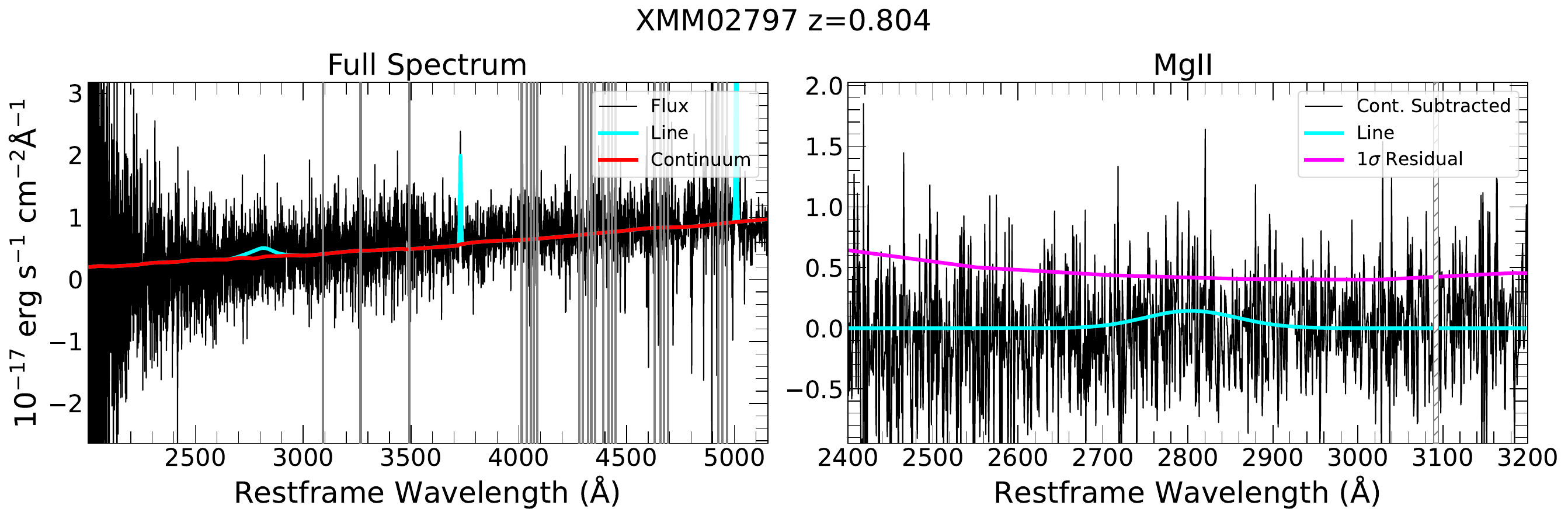}
    \caption{Example fitting of AGN with SDSS spectra. The top panel shows a type-1 AGN with clear MgII detection while the bottom AGN shows the spectra of an AGN with no MgII detection. The left panel shows the full restframe spectra while the right panel shows a zoomed-in on the position of the MgII emission line. The best-fitted pseudo-continuum model is shown with the red line while the emission lines are shown in cyan. The magenta line in the right panel shows the 1$\sigma$ residuals. Hatched regions are wavelengths that are affected by strong sky emission lines of telluric absorption. The grey-shaded regions mark the wavelength affected by strong sky emission lines. }
    \label{fig:examplespec}
\end{figure*}

Figure \ref{fig:examplespec} shows the fitting results of an AGN with a clear detection of MgII and an AGN with no clear broad MgII emission lines. Among the 239 AGN with SDSS spectrum, 92 AGN require 2 Gaussian components, while 124 AGN require a single Gaussian. 23 AGN have no broad emission lines based on visual examination. The 23 AGN with no broad emission lines were classified as 17 narrow-line AGN and 6 AGN with low-S/N spectra. The narrow-line AGN generally show a reddened continuum with no obvious broad lines. The spectroscopic redshift was determined from the [OII] emission lines at $3737$\AA \  and is consistent with the redshift catalogued in the SDSS database. In some cases, [OIII] emission lines and weak narrow H$\beta$ were also detected. For AGN with low-S/N spectra, reliable spectroscopic redshift can not be determined, nor can broad emission lines be confirmed. The spectroscopic redshift previously adopted from SDSS was replaced with the photometric redshift instead. 

To estimate the uncertainty, 25 mock spectra were constructed for each AGN. Each pixel was drawn from a normal distribution centred on the best-fitted flux and has a dispersion according to the 1$\sigma$ photometric uncertainties. Then the fitting was performed again for each realisation. For the FWHM, luminosity, and S/N, we adopted the median value and calculated the lower and upper uncertainty from the 16th and 84th percentile of the mock distribution.

\begin{figure}
    \centering
    \includegraphics[width=\linewidth]{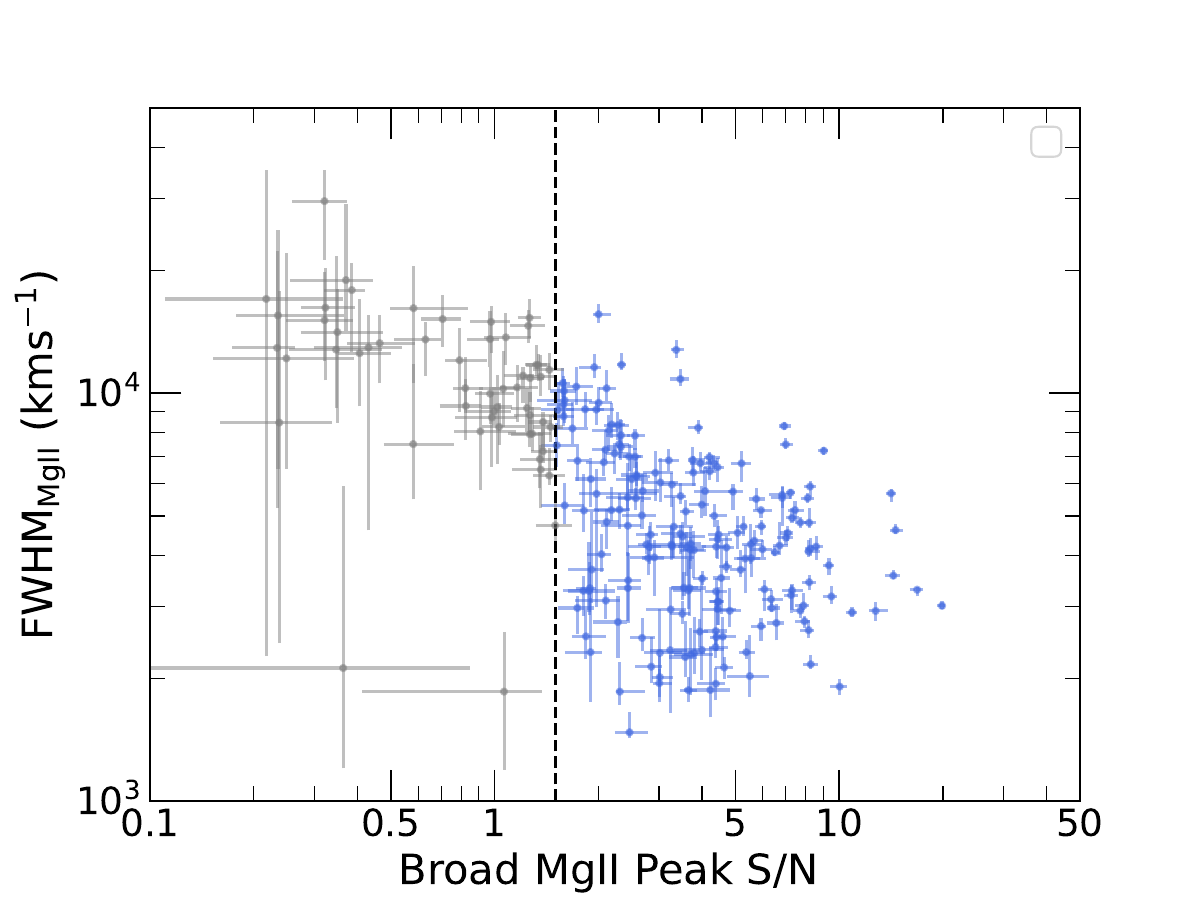}
    \caption{Peak signal-to-noise ratio against the FWHM of the broad MgII emission line component. Blue data points are AGN with detected MgII emission lines.}
    \label{fig:mg2sn}
\end{figure}

The peak signal-to-noise ratio is defined as the ratio between the peak flux and the noise level at the position. Figure \ref{fig:mg2sn} shows the peak S/N of the broad component of the MgII emission line against the MgII FWHM. Spectra with low peak S/N have large MgII FWHM due to continuum noise and not from the possible emission line.  The  MgII emission line was considered as detected if the peak S/N is larger than 1.5, 190 broad-line AGN from SDSS satisfy this criterion.

\begin{figure}
    \centering
    \includegraphics[width=\columnwidth]{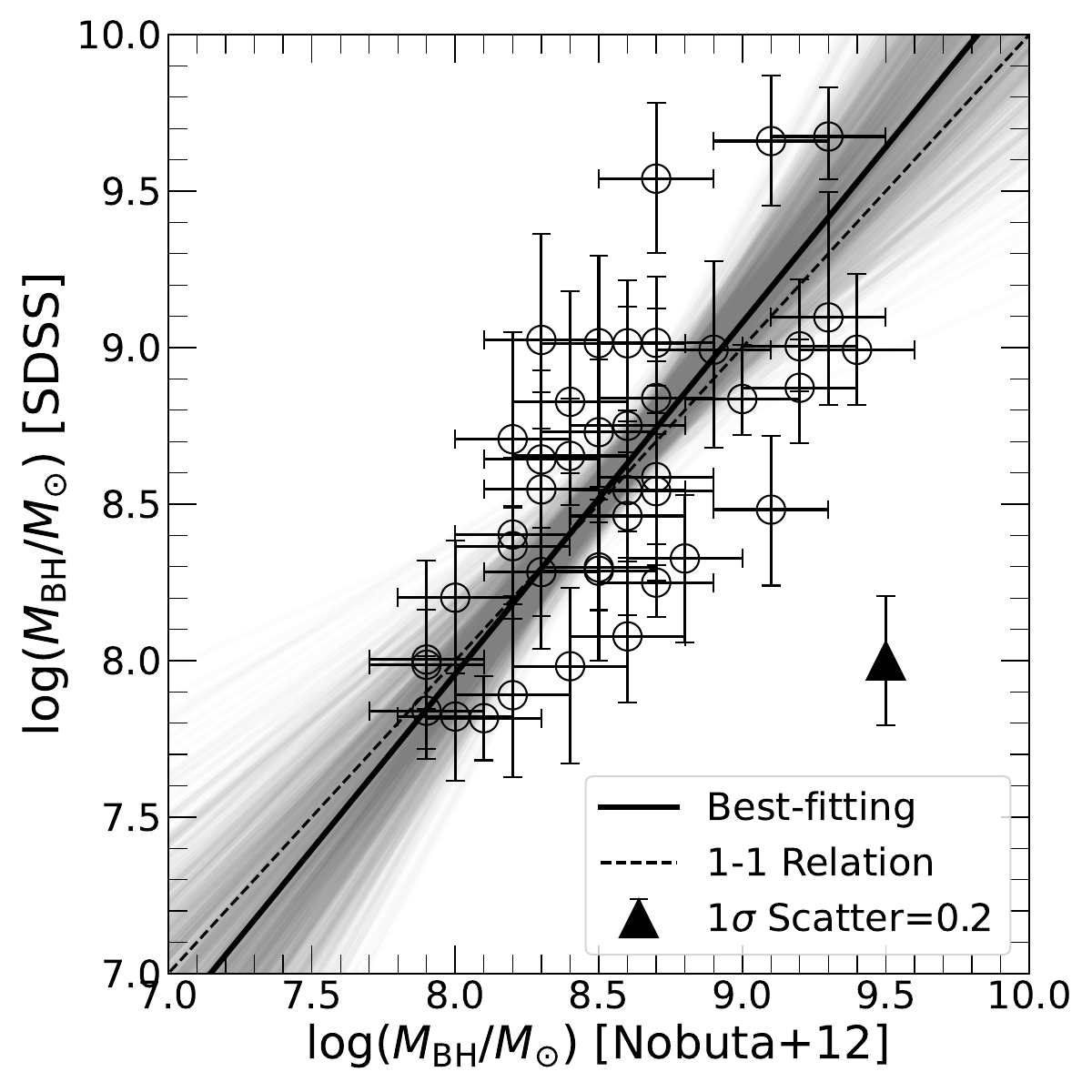}
    \caption{Comparison between the black hole mass estimated by \citet{2012ApJ...761..143N} from FOCAS spectra against our measurements using SDSS spectra. The best-fitted relation is shown with a thick black line. The dashed line shows the 1-1 relation. The filled triangle shows the $1\sigma$ scatter of the data points.}
    \label{fig:mbhcomp}
\end{figure}

We used the single-epoch virial mass relationship of \citet{2009ApJ...699..800V} to infer the black hole mass of the broad-line AGN using the MgII FWHM and the continuum luminosity at 3000\AA. The relationship is expressed as 

\begin{equation}
    M_{\rm BH} {\rm [M_{\odot}]}=10^{6.86}\Bigg[\frac{{\rm FWHM}_{\rm MgII}}{1000\ {\rm km\ s^{-1}}}\Bigg]^2 \Bigg[\frac{\lambda_{\rm 3000} L_{\rm 3000}}{10^{44}\ {\rm erg\ s^{-1}}}\Bigg]^{0.5}
    \label{eq:masseq}
\end{equation}

where $\rm FWHM_{MgII}$ is the FWHM of the MgII emission line and $L_{\rm 3000}$ is the continuum luminosity density at 3000 \AA. We estimated the uncertainty of the SMBH mass estimate by propagating the FWHM and $L_{\rm 3000}$ uncertainty through equation \ref{eq:masseq} with an additional 0.1 dex uncertainty from the comparison with $
\rm H\beta$ and CIV mass estimates \citep{2009ApJ...699..800V}. In addition to the SDSS broad-line AGN, we added 109 SMBH mass measurements from \cite{2012ApJ...761..143N}, who estimated the SMBH mass of broad-line AGN within the SXDS survey region \citep{2004AAS...205.8105S,2015PASJ...67...82A}. \citet{2012ApJ...761..143N} also used the virial mass relationship of \citet{2009ApJ...699..800V} to estimate the SMBH mass of MgII detected AGN observed with the FOCAS spectrograph \citep{2002PASJ...54..819K}. For a sub-sample of AGN observed in the near-infrared with the FMOS spectrograph \citep{2010PASJ...62.1135K}, they also determined SMBH mass using the $\rm H \alpha$ emission line, which is in good agreement with that determined with MgII emission line. \citet{2012ApJ...761..143N} did not present the uncertainty for each AGN, therefore, we adopted a black hole mass uncertainty of 0.2 dex for their estimate based on their discussion of the black hole mass uncertainty which includes the uncertainty from the FWHM, continuum luminosity, and FeII template. We evaluated the consistency within the sample by performing linear regression using 42 AGN that has black hole mass from \cite{2012ApJ...761..143N} and from our analysis using the SDSS spectra as shown in Figure \ref{fig:mbhcomp}.  The results suggested that the black hole mass is statistically consistent with each other within $1\sigma$ scatter ($\sigma_{\rm scatter}=0.2$) of the data. We adopted the measurements from \citet{2012ApJ...761..143N}, considering the higher S/N of their spectroscopic data. The final number of black hole mass estimations are 106 from \citet{2012ApJ...761..143N} and 148 from SDSS. 

\subsection{SED Fitting} \label{sec:analy:sedfit}
We performed SED fitting using CIGALE 2022 \citep{2019A&A...622A.103B,2020MNRAS.491..740Y} to estimate the host galaxy's stellar mass and the total star-formation rate (SFR). CIGALE is a SED fitting code that balances the UV radiation absorbed by dust and re-emitted in the infrared. The parameters on the host galaxy properties were computed using Bayesian posterior distributions.  

We used the delayed star-formation history and the Chabrier initial mass function \citep{2003PASP..115..763C}.  For the dust model, we adopted the empirical dust emission SEDs from \citet{2014ApJ...784...83D}, which contains only two parameters. The choice of the simpler dust model is due to the limited number of FIR detections in the dataset. Only upper limits are available for the majority of the sample and even  FIR-detected AGN are detected only in one or two bands of \textit{Herschel}/SPIRE with S/N less than 10. We used the SKIRTOR AGN models \citep{2012MNRAS.420.2756S,2016MNRAS.458.2288S} to fit the mid-infrared and optical/UV AGN continuum. We do not use the X-ray module to avoid introducing additional systematics through the absorption correction of the X-ray fluxes using hardness ratio which may affect the optical/mid-infrared models. Moreover, several sources have only upper or lower limits on the extinction correction due to detection in only a single band.  We described this in detail in section \ref{sec:analy:nhfhr}.   The parameters used for the SED fitting are presented in table \ref{tab:sedfittingparam}.  

The fitting was performed separately for the type-1 and type-2 AGN. For broad-line AGN and the morphologically selected type-1 AGN, the inclination angle of the SKIRTOR AGN model was set to 30 degrees (face-on) and the AGN fraction ($f_{\rm AGN}$) which is calculated between 0.1-1\um was set to $f_{\rm AGN}>0.5$. By doing so, the AGN continuum emission is forced to dominate the optical/UV continuum. In type-2 AGN, the optical/UV is dominated by the host galaxy's stellar emission but the AGN dominates the mid-infrared emission. Therefore, the inclination angle was set to 70 degrees instead (edge-on) and the AGN fraction was calculated between 2-10 \um with $f_{\rm AGN}>0.5$ instead. 

\begin{figure*}
    \centering
    \includegraphics[width=\columnwidth]{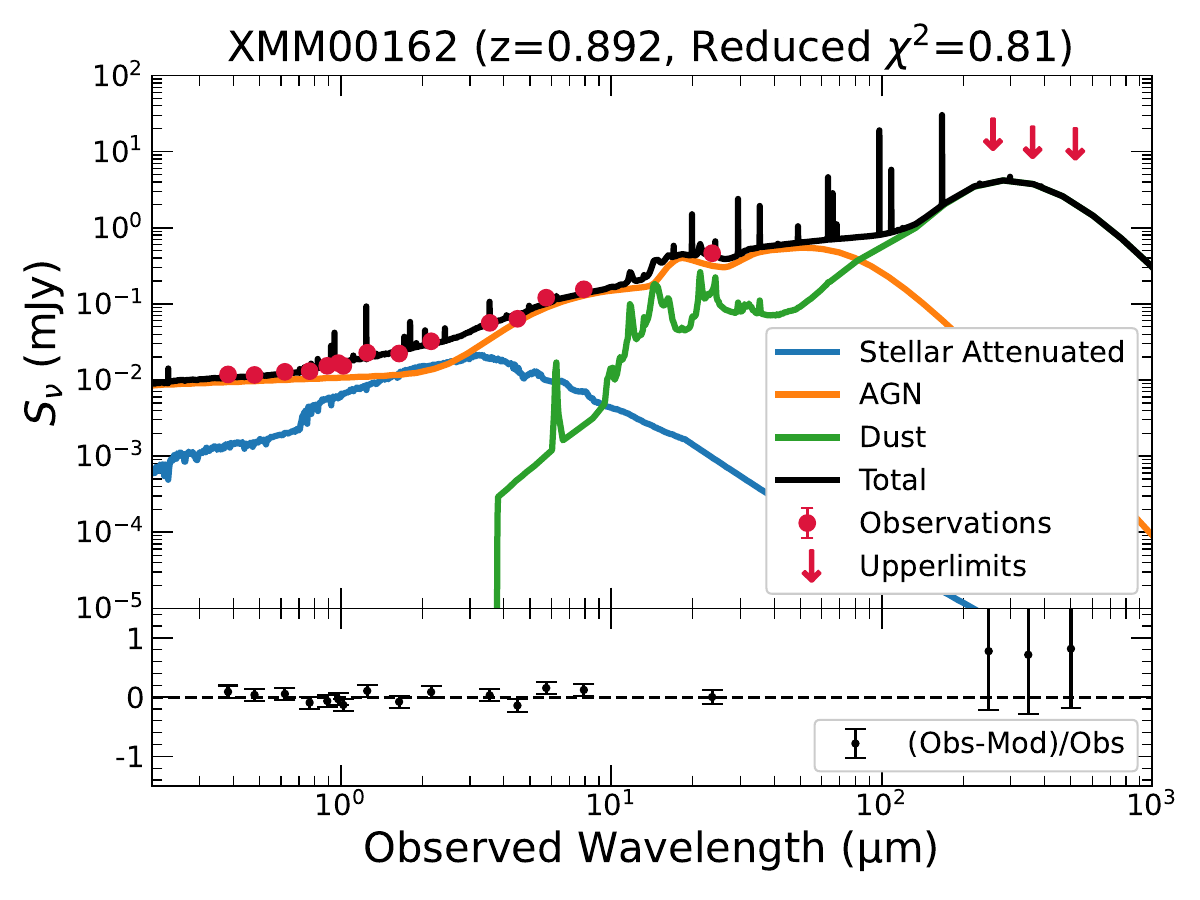}
    \includegraphics[width=\columnwidth]{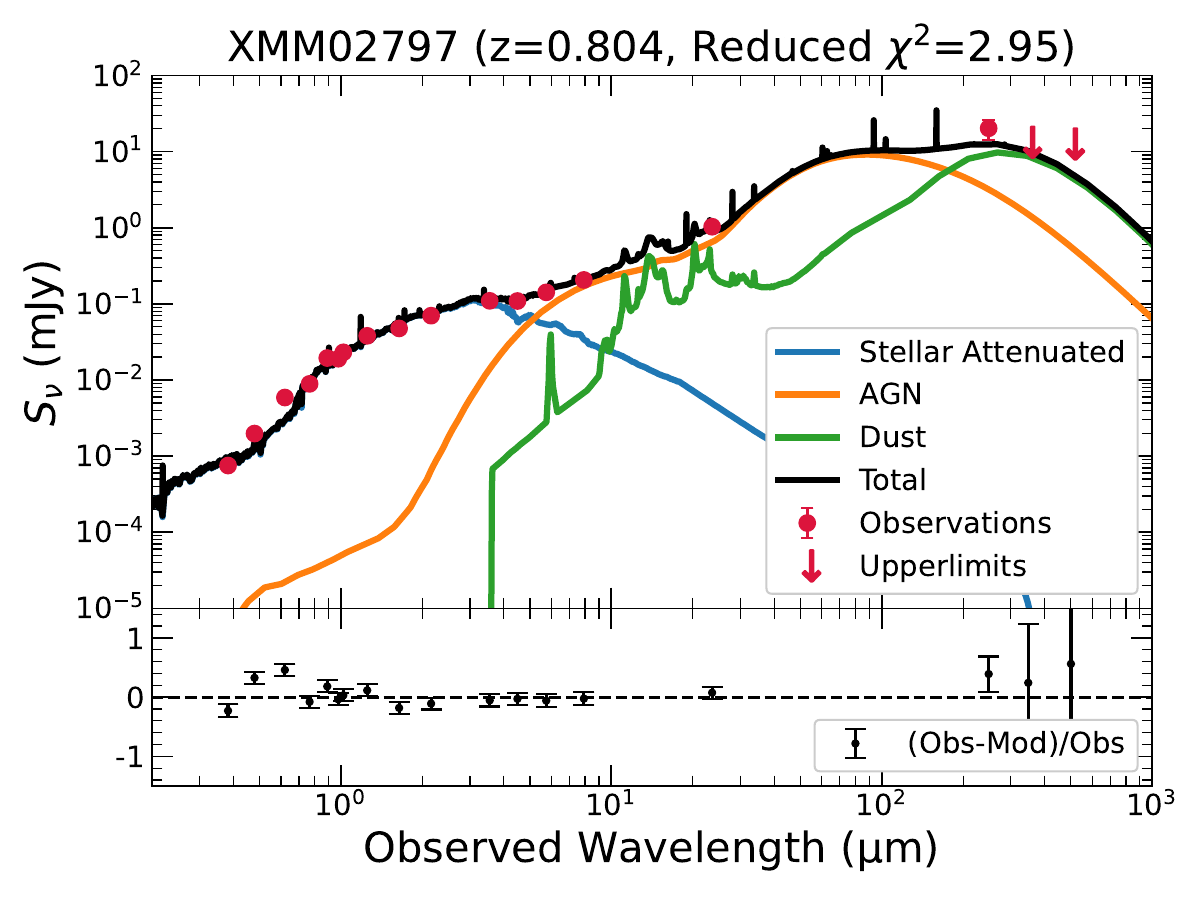}
    \caption{The best-fitted observed SED of the type-1 and type-2 on the left and right, respectively. The top panels show the model fitting while the bottom panels show the relative uncertainty. The observations are shown with red datapoint while upper limits are shown with downward arrows. The attenuated stellar, AGN, dust, and total models show blue, orange, green, and black respectively.}
    \label{fig:examplesed}
\end{figure*}

Figure \ref{fig:examplesed} shows the best-fitted SED of the type-1 and type-2 AGN presented in figure \ref{fig:examplespec}. To have reliable stellar mass and total SFR measurements, we required low reduced $\chi^2$ values that indicate good SED fits. Thus, we applied a threshold of reduced $\chi^2=5$ that has been adopted in previous studies (\citet{2022A&A...667A..56P}, and references therein)as well as by visual examination of the best-fitted models to exclude sources with bad fits. The median reduced $\chi^2$ of the type-1 and type-2 AGN is 2.2 and 1.7 with 87\% and 92\% of each sample with a reduced $\chi^2<5$, respectively.

\begin{table*}
	\centering
	\caption{CIGALE for type-1 AGN. Parameter values in parentheses indicate the values used for type-2 AGN.}
	\label{tab:sedfittingparam}
	\begin{tabular}{lc} 
	    \hline
        \multicolumn{2}{c}{Star-formation History (sfhdelayed)}  \\
        \hline
            $e$-folding time of main population ($\tau_{\rm Main}$) & 100,500,1000,2000,4000,6000,8000 Myr \\
            Age of the main stellar population ($\rm Age_{\rm Main}$) & 1000,2000,3000,4000,6000,8000,10000 Myr \\ 
            $e$-folding time of the late starburst population ($\tau_{\rm Burst}$) & 3 \\
            Age of the late burst in Myr ($\rm Age_{\rm Burst}$) & 100.0 \\
            Mass fraction of the late burst population ($\rm f_{\rm Burst}$) & 0.0,0.01 \\ 		
        \hline
        \multicolumn{2}{c}{SSP Model (BC03)}  \\
        \hline
            Initial mass function (IMF) & Chabrier \\ 
            Metallicity & 0.02 \\
            Separation Age & 10.0 Myr \\
        \hline
        \multicolumn{2}{c}{Dust Attenuation (dustatt\_modified\_starburst)}  \\
        \hline
            The colour excess of the nebular lines (E(B-V)) & 0,0.2,0.4,0.6,0.8,1.0,1.5,2.0 \\
            Stellar E(B-V) conversion factor & 0.44 \\
            Ext law emission lines & \citet{1989ApJ...345..245C} \\
            Ratio of total to selective extinction ($\rm R_V$) & 3.1 \\
        \hline
        \multicolumn{2}{c}{Dust Emission \citep{2014ApJ...784...83D}}  \\
        \hline
            $\rm f_{AGN}$ & 0 \\
            Alpha slope ($\alpha$) & 0.0625,1.0,2.0,3.0,4.0 \\
        \hline
        \multicolumn{2}{c}{AGN Model (SKIRTOR 2016)}  \\
        \hline 
            Average edge-on optical depth at 9.7$\rm \mu m$ ($\tau$) & 3,7 \\
            Power-law exponent ($pl$) & 0.5 \\
            Dust density gradient index ($q$) & 0.5 \\
            Torus equatorial plane and edge angle ($oa$) & 40 Deg \\
            The ratio of torus outer to the inner radius ($R$) & 20 \\ 
            Inclination angle ($i$) & 30(70) Deg \\
            Optical power-law of index ($\delta$) & -0.36 \\
            AGN fraction $\rm f_{\rm AGN}$ & 0.5,0.6,0.7,0.8,0.99 \\
            AGN fraction band  & 0.1/1.0 (2.0/10.0) $\rm \mu m$\\
            Extinction law & SMC \\
            Polar dust E(B-V) &  0.0,0.25,0.5,0.75,1,2 \\
            Polar dust temperature & 100 Kelvin\\
            Emissivity index of the polar dust & 1.6 \\
        \hline	\end{tabular}
\end{table*}

\subsection{Estimation of the Star-formation Rates} \label{sec:analy:sfrest}

To evaluate the SFR properties of the host galaxies, i.e. location in the UVJ and star-formation main sequence diagrams, without the contamination from AGN emission, we used the best-fitted models from CIGALE with all AGN components removed for the calculation. Firstly, we calculated the restframe $U-V$, and $V-J$ band colours assuming the Subaru U and V filters and the UKIRT J-band filters. Secondly, we calculated the total SFR rate and specific star-formation rate (SSFR) from the host galaxy optical and far infrared emission using the relationship presented by \citet{2011ApJ...737...67M} which is given as $$\frac{\rm SFR_{\rm tot}}{M_{\odot} \rm yr^{-1}} =4.42\times 10^{-44}\bigg(\frac{L_{\rm FUV}+0.88L_{IR}}{\rm erg\ s^{-1}} \bigg)$$ where $L_{\rm FUV}$ is the FUV luminosity assuming the GALEX FUV transmission curve and $L_{\rm IR}$ is the infrared 8-1000\um luminosity. We calculated $L_{\rm FUV}$ and $L_{\rm IR}$ from the best-fitted model without the AGN component. CIGALE also provides an estimate of the instantaneous SFR based on the assumed star formation history. We checked the consistency between the instantaneous SFR (sfh.sfr) from CIGALE  and compared it with the SFR from the FUV and IR luminosity. The two are consistent with each other within the 1$\sigma$ uncertainties.  We adopted the total SFR from the FUV and infrared luminosity to escape from the dependence on the assumption of the star formation history.

CIGALE provides the uncertainty for the total (AGN and host galaxy) FUV luminosity and the total dust emission. To estimate the SFR uncertainty, we assumed that the uncertainty of the best-fit AGN FUV luminosity is the same as the total FUV luminosity when the AGN components are removed and assume that the relative uncertainty of the 8-1000\um is the same as the relative uncertainty of the total dust luminosity. The uncertainties are then propagated through the SFR relationship. 

\subsection{$N_{\rm H}$ and 2-10 keV AGN luminosity} \label{sec:analy:nhfhr}

In this work, the hydrogen column density  ($\log N_{\rm H}$) was determined from the X-ray hardness ratios assuming an X-ray spectral model and the redshift. The X-ray hardness ratio (HR) is defined as $HR=(H-S)/(H+S)$ where S is the count rates in the 0.5-2 keV band and H is the count rates in the 2-10 keV band. Here, we assumed the same phenomenological AGN model used in \citet{2022ApJ...941...97V} to calculate the fluxes. The model includes an exponentially declining power-law AGN continuum component, a Thompson-scattered AGN component, and a Compton-reflection component \citep{1995MNRAS.273..837M}. The scattered component is set to 1\% of the power-law continuum with the same photon index while the reflection component strength was set to $rel_{\rm Refl}$=1, equal to the continuum. Photoelectric absorption (tbabs) was applied only to the power-law continuum. The count-rate calculations were performed using XSPEC \citep{1996ASPC..101...17A} assuming an on-axis response matrix file (RMF) and ancillary matrix files (ARF) of the {\it XMM-Newton} PN detector.\footnote{https://www.cosmos.esa.int/web/xmm-newton/epic-response-files}

\begin{figure*}
    \centering
    \includegraphics[width=\linewidth]{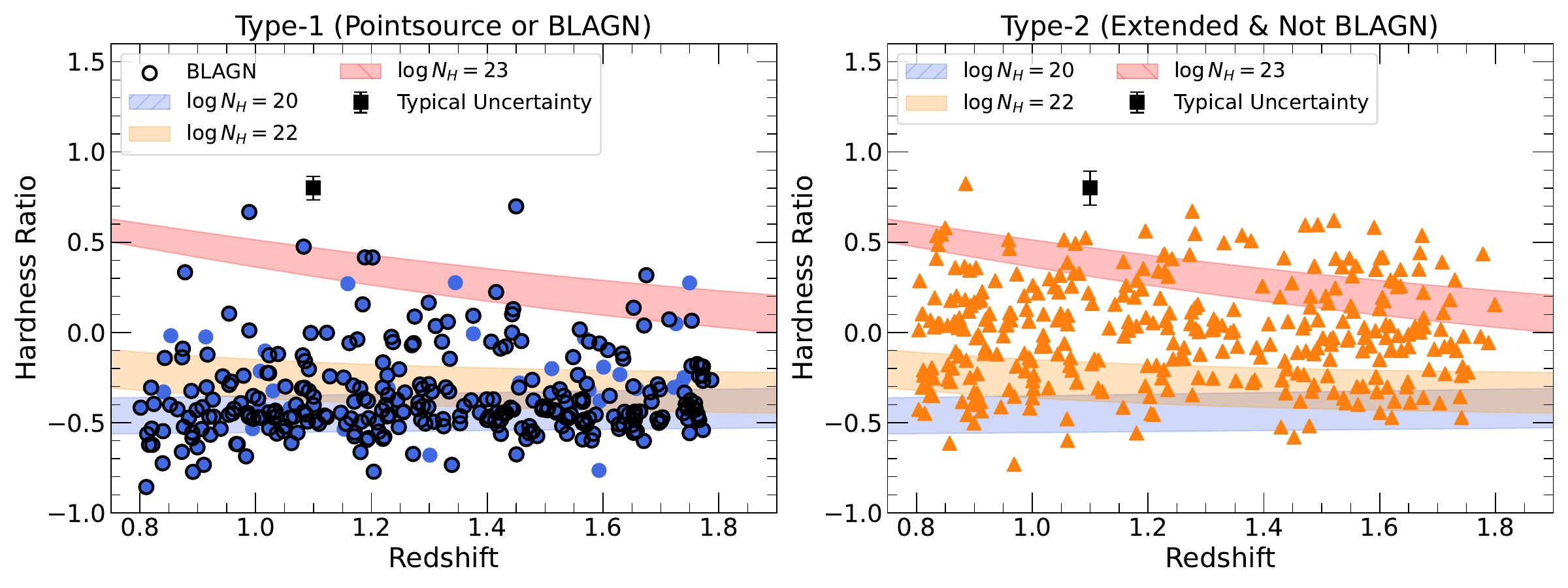}
    \caption{The X-ray hardness ratios of type-1 AGN on the left panel and type-2 AGN on the right panel are shown with blue round and orange triangle symbols, respectively. Spectroscopically confirmed broad-line AGN are shown with black-rimmed symbols. The black square symbol shows the typical uncertainty in the HR measurement of the AGN in each sample. The blue, yellow, and red bands show the hardness ratio tracks assuming $\log N_{\rm H}\ [{\rm cm^{-2}}]$=20,22, and 23. The bottom edge to the top edge of each band shows the HR tracks assuming a photon index of 2.0 to 1.5.}
    \label{fig:zhardness}
\end{figure*}

Figure \ref{fig:zhardness}  shows the X-ray HR of type-1 and type-2 AGN. The X-ray HR is a function of the intrinsic column density, power-law index, and redshift as shown with the three bands. The upper and lower edge of each band represent the HR assuming a photon index of 1.5 and 2.0 which is the typical range of the photon index observed in unobscured AGN, with a nominal value of 1.8 \citep{2017ApJS..233...17R}. An increase in the Hydrogen column density will reduce the count rate in the 0.5-2 keV band more than in the 2-10 keV band. This increases the HR with increasing hydrogen column density. For high redshift AGN, hard X-rays which are less susceptible to absorption are shifted into the softer bands and reduce the HR  with increasing redshift. As a result, the observed HR is a degenerate parameter between the photon index and the column density. The same HR can be reproduced with either an AGN with a large photon index (steep) with a larger column density or a smaller (flat) photon index with a smaller column density. We included this systematic effect in the uncertainty of the hydrogen column density.{\color{red} \textbf{}}

\begin{figure*}
    \centering
    \includegraphics[width=\linewidth]{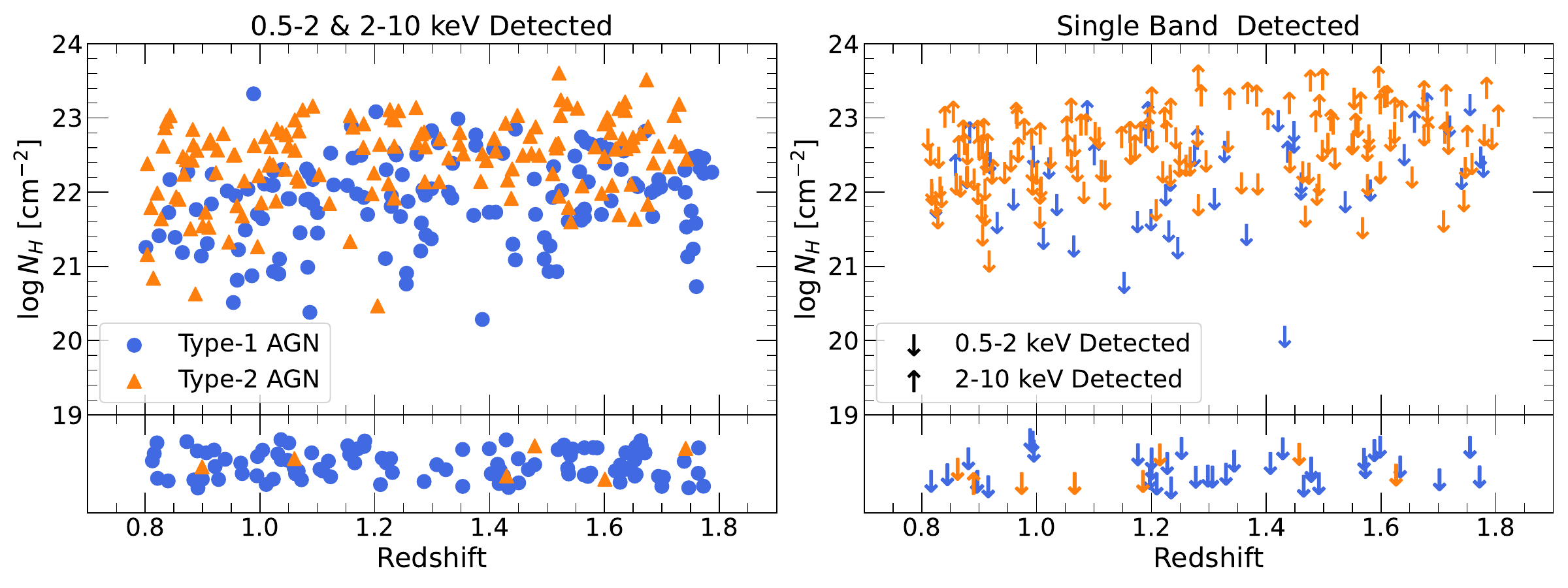}
    \caption{The hydrogen column density of AGN detected in both 0.5-2 and 2-10 keV bands (left) and single band detected AGN (right). The top panel shows the AGN which has HR within the model grid. The bottom panel shows the AGN which has HR softer than the model grid and is set to have $\log N_{\rm H}=20$. Type-1 AGN are shown with blue round symbols while type-2 AGN are shown with orange triangle symbols.  For single band detected AGN, lower limits are shown with up arrows while upper limits are shown with down arrows. }
    \label{fig:znh}
\end{figure*}

Figure \ref{fig:znh} shows the hydrogen column density determined from the X-ray HR assuming a photon index of 1.8. On the left panel, we show the sample of AGN which is detected in both the  0.5-2 keV band and the 2-10 keV band while the right-hand panel shows the sample of AGN which is detected only in a single energy band. The bottom part of each panel shows the AGN that has an X-ray HR softer than a power law with a photon index of 1.8. For these AGN, the hydrogen column density was set to \lognh{=20}.  For objects that are detected only in the 0.5-2 keV band (down arrows), only upper limits on the hydrogen column density can be determined.  On the other hand, for objects that are detected only in the 2-10 keV band (up arrows), only lower limits on the hydrogen column density can be determined. For AGN which is detected only in the 0.5-10 keV band, no constraints on the hydrogen column density can be made. Table \ref{tab:xray_stats} shows the number of unobscured (\lognh{<22}) and obscured (\lognh{\geq22}) AGN among the type-1 and type-2 AGN, separated based on the detection band.

\begin{table*}
    \centering
    \begin{tabular}{l|c|c|c|c|c|c|c|c|c}
         \hline
         Detection & \multicolumn{2}{c}{0.5-2 \& 2-10 keV} & \multicolumn{2}{c}{Only 2-10 keV} & \multicolumn{2}{c}{Only 0.5-2 keV} & \multicolumn{2}{c}{Only 0.5-10 keV} \\
         \lognh{} & $<22$ & $\geq22$ & $<22$ & $\geq22$ & $<22$ & $\geq22$ & \multicolumn{2}{c}{
         Unconstrained} & Total \\
         \hline
         Type-1 & 175 & 67 & 0 & 10 & 52 & 19 & \multicolumn{2}{c}{5} & 328 \\
         Type-2 & 37 & 116 & 0 & 62 & 27 & 73 & \multicolumn{2}{c}{23} & 338 \\
         \hline
         Total &  212 & 183 & 0 & 72 & 79 & 92 & \multicolumn{2}{c}{28} & 666 \\
         \hline
    \end{tabular}
    \caption{Number of unobscured (\lognh{<22}) and obscured (\lognh{\geq22}) AGN among the type-1 and type-2 AGN, separated based on the detection bands. }
    \label{tab:xray_stats}
\end{table*}

\begin{figure*}
    \centering    
    \includegraphics[width=\linewidth]{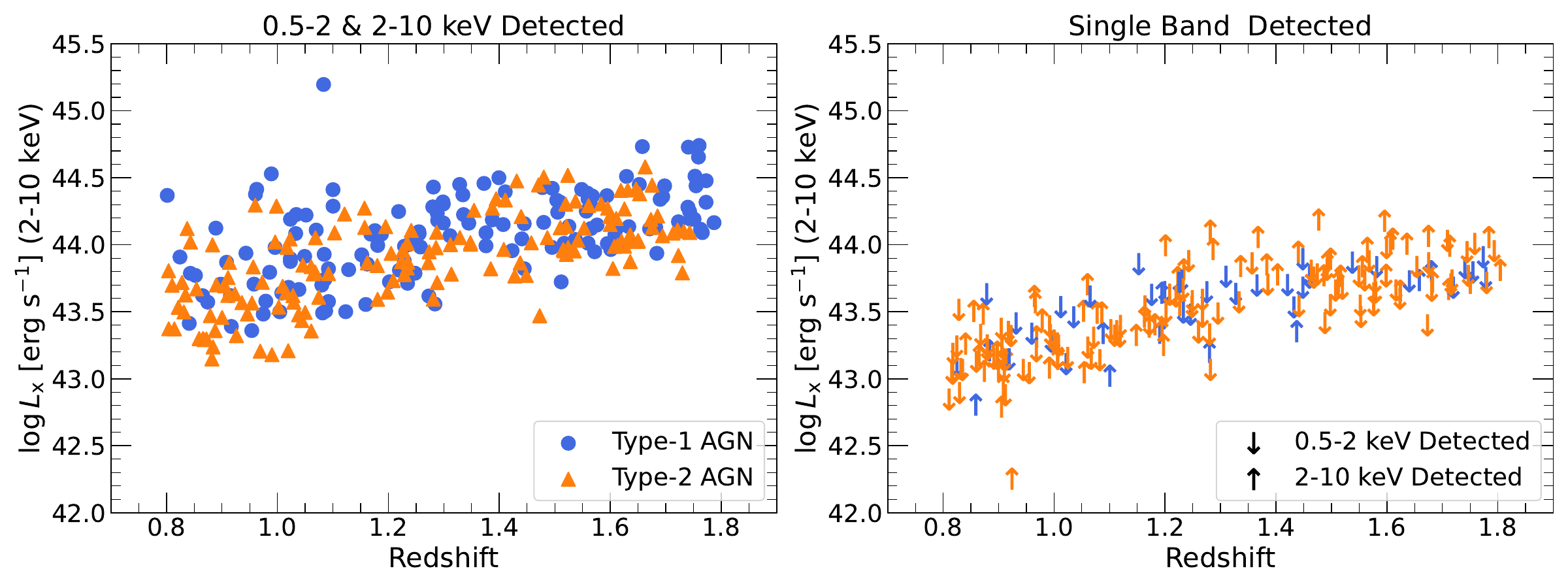}
    \caption{The 2-10 keV luminosity of AGN detected in both 0.5-2 and 2-10 keV bands (left) and single band detected AGN (right).  Type-1 AGN are shown with blue round symbols while type-2 AGN are shown with orange triangle symbols. For single band detected AGN, lower limits are shown with up arrows while upper limits are shown with down arrows.  }
    \label{fig:zlx}
\end{figure*}

 The flux presented \citet{2018MNRAS.478.2132C} is the weighted average of the flux measured in the PN, MOS1, and MOS2 detectors. The flux in each detector was converted from the count rates using a power-law model with a photon index ($\Gamma$) of 1.7.  Here, we converted the flux back to ``PN-equivalent'' count rates using the energy conversion factor (ECF) presented in \citet{2018MNRAS.478.2132C} and then recalculated the flux assuming a photon index of 1.8 instead. The ECF in soft-band, hard-band, and full-band is $6.84\times 10^{11}, 1.26\times 10^{11}$, and $3.36 \times 10^{11}\ \rm count\ s^{-1}\ / erg\ s^{-1} cm^{-2}$.  Using the best available redshift and hydrogen column density,  the intrinsic 2-10 keV X-ray AGN luminosity corrected for X-ray absorption ($L_X$) was calculated using

$$L_X ={\rm ext}(z,N_{\rm H})  k(z)  4\pi D_L^2(z)  f_X$$

where $f_X$ is the X-ray flux in either soft-band, hard-band, or full-band. $ {\rm ext}(z,N_{\rm H})$ is the absorption correction which is a function of redshift and hydrogen column density, $k(z)$ is the X-ray k-correction factor based on the unobscured AGN spectra, and $D_L$ is the luminosity distance. Figure \ref{fig:zlx} shows the 2-10 keV luminosity as a function of redshift for AGN detected in both 0.5-2 keV band and 2-10 keV band and for AGN detected in only a single band.

To evaluate the reliability of the  SED fitting results for the AGN continuum, we examined the relationship between the X-ray 2-10 keV luminosity and the 6\um AGN luminosity. The two luminosities correlate with each other over 3 orders of magnitude \citep{2009ApJ...693..447F,2015ApJ...807..129S,2019MNRAS.484..196T,2022A&A...661A..15T}. Observations of high luminosity AGN ($\log \lambda L_{6\mu m}\ {\rm [erg\ s^{-1}]}>46$) suggested that X-ray emission becomes weak at high luminosity and high Eddington ratio systems due to disruption of the corona by radiatively inefficient accretion flows or shielding from winds \citep{2019A&A...632A.109N,2020A&A...635L...5Z,2020A&A...636A..73D}. Therefore above $\log \lambda  L_{6\mu m}\ {\rm [erg\ s^{-1}]}>46$, the $L_X-L_{6\mu m}$ relationship deviates from linear relationships. We performed a linear regression analysis using the Bayesian linear regression routine\footnote{\url{https://github.com/jmeyers314/linmix}} presented by \citet{2007ApJ...665.1489K}.  Since the X-ray model CIGALE was not used in the fitting, the measurements of the X-ray luminosity and the infrared luminosity are independent of each other. The lower and upper uncertainties of the linear regression coefficients were estimated based on the 16th and 84th percentiles of 10,000 mock-fitting results. 

\begin{figure}
    \centering
    \includegraphics[width=\columnwidth]{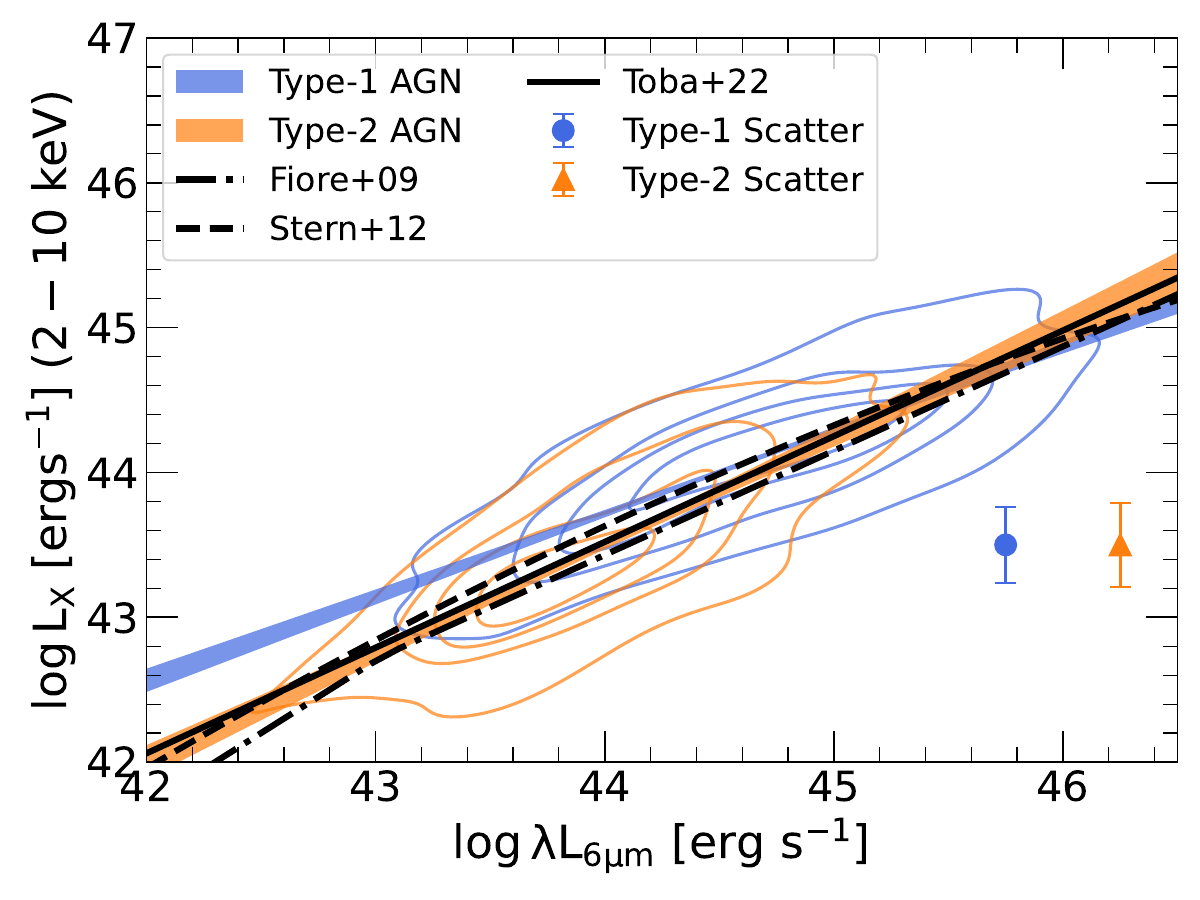}
    \caption{The best-fitted $L_{6\rm \mu m}-L_{\rm X}$ relationship of type-1 and type-2 AGN shown in blue and orange bands, respectively. The width of each band represents the $1\sigma$ confidence interval. The distribution of type-1 and type-2 AGN are shown in blue and orange contours. The blue round and orange triangle data point shows the $1\sigma$ scatter of the type-1 and type-2 AGN samples. The best-fitted relationships are compared to the results from \citet{2009ApJ...693..447F,2015ApJ...807..129S} and \citet{2022A&A...661A..15T}}
    \label{fig:l6lx}
\end{figure}

Figure \ref{fig:l6lx} shows the 2-10 keV X-ray luminosity vs. the 6\um AGN luminosity relationship of the 2-10 keV detected sample of AGN with spectroscopic redshift. We used the 2-10 keV detected AGN because it is less affected by absorption compared to the other bands. Here, the redshift range was not limited to 0.8-1.8 since most type-2 AGN with spectroscopic redshift is at $z<0.8$. \loglx{=43.5-44.5} which corresponds to the luminosity of our sample of redshift 0.8-1.8 AGNs, our results are consistent with all of the relationships determined by previous studies within the 1$\sigma$ uncertainty and the scatter.  In addition, the relationship is still consistent with a linear relationship since the sample of AGN used to perform the analysis does not contain highly luminous AGN. Equation \ref{eq:lxlir} shows the best-fit linear relationship between the 6\um infrared AGN luminosity and 2-10 keV AGN luminosity based on the hard-band selected sample.

\begin{equation}
    \log L_{\rm X}=0.58_{-0.03}^{+0.02} \log L_{\rm 6 \mu m} + 18.39_{-0.25}^{+0.28}, \sigma_{\rm scat}=0.26
    \label{eq:lxlir}
\end{equation}

where $L_{\rm X}$ is the 2-10 keV AGN luminosity and $L_{\rm 6 \mu m}$ is the 6 \um AGN luminosity. The 1$\sigma$ scatter ($\sigma_{\rm scat}$) is 0.26.

\section{Results}
\subsection{Host Galaxy Stellar Mass} \label{sec:res:mstellar}

\begin{figure}
    \centering
    \includegraphics[width=\columnwidth]{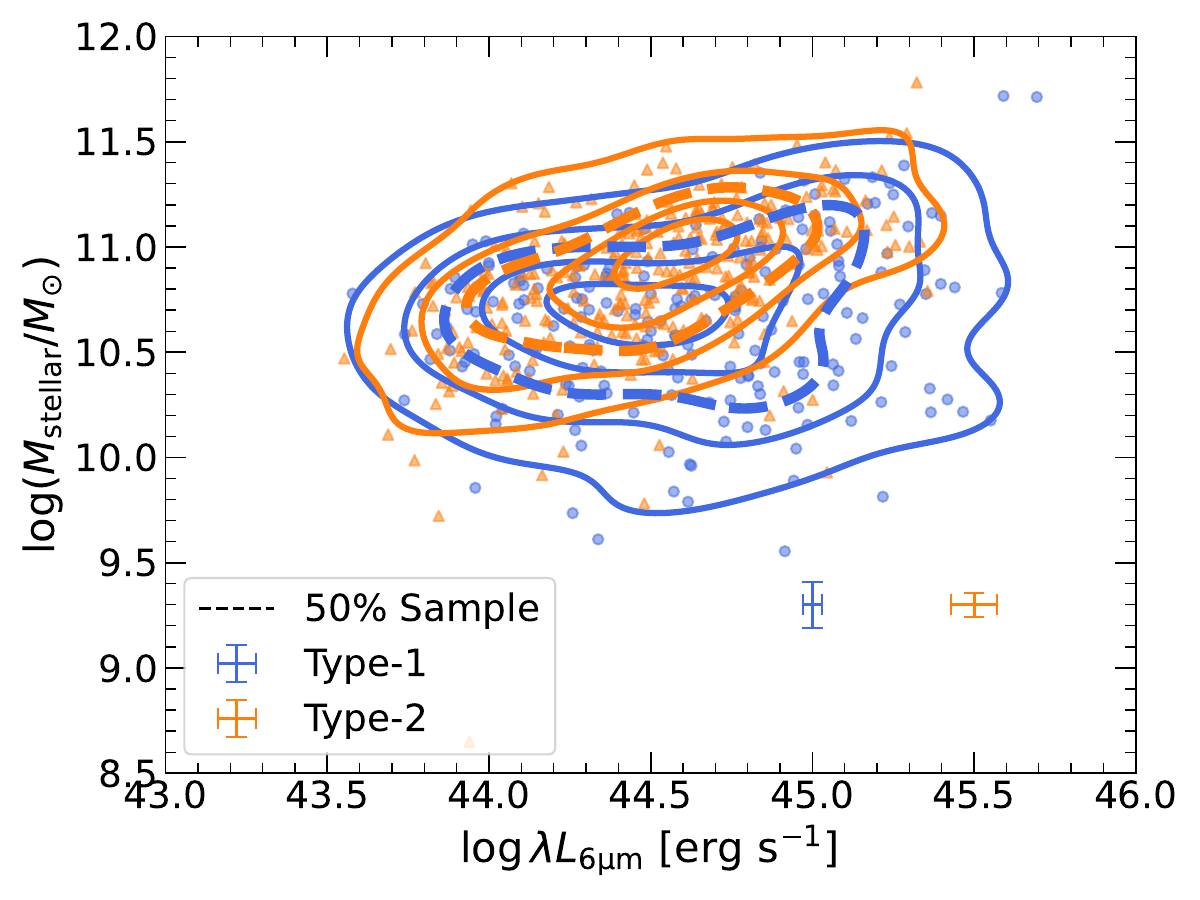}
    \caption{The 6\um luminosity against stellar mass $M_{\rm stellar}$ of type-1 AGN and type-2 AGN shown as blue and orange contours, respectively. The contour levels show the area which 10-90\% of the datasets lay under. The 50\% level is shown with dashed contours. The faceless data point with error bars shows the typical uncertainties of the AGN.}
    \label{fig:l6mstellar}
\end{figure}

First, we examined the host galaxy stellar mass of type-1 and type-2 AGN  which were estimated from the SED fitting. In this section as well as in section \ref{sec:res:mbh} and section \ref{sec:res:sfa}, we limited the sample to 497 AGN detected in either the 0.5-2 keV band or 2-10 keV band and have a reduced $\chi^2<5$. Furthermore, for type-1 AGN, we use only those with black hole mass from spectroscopically confirmed
broad lines. This includes 205 type-1 AGN and 292 type-2 AGN. Figure \ref{fig:l6mstellar} shows the stellar mass and 6-micron AGN luminosity of type-1 and type-2 AGN. Type-1 AGN and type-2 AGN in our sample have an average host galaxy stellar mass of $\log M_{\rm stellar}=10.7_{-0.3}^{+0.4}$ and  $\log M_{\rm stellar}=10.9_{-0.3}^{+0.3}$, respectively. The stellar mass uncertainty of type-1 AGN is on average a factor of 2 larger than type-2 AGN. This is due to contamination by the AGN optical emission among type-1 AGN which results in less certain stellar mass.  The AGN emission can be fitted with a younger stellar population with a lower mass-to-light ratio or an SED with a larger burst fraction. The sample currently has no reliable stellar age indicators from spectral analysis but the stellar age of the best-fitted SED does not show any strong difference between type-1 and type-2 AGN and the burst fraction used in our fitting is limited to at most 1\% of the total stellar mass. We further examined the best-fitted host galaxy model and calculated the best-fit H-band mass-to-light ratio. The H-band rest-frame luminosity was calculated by applying the H-band filter to the best-fit SED. We found that type-1 AGN have a mass-to-light ratio lower than type-2 AGN by approximately 0.1 dex but within the scatter of the mass-to-light ratio of both type-1 AGN and type-2 AGN. We added a systematic uncertainty of 0.1 to the stellar mass, star formation rate, and specific star formation rate of type-1 AGN in the quadrature to account for the systematic difference in the mass-to-light ratio.

The sample covers a broad range in redshift and luminosity and can be affected by the cosmological increase of luminous AGN towards redshift $\sim 2$. In addition, the AGN sample contains a mixture of AGN detected in 0.5-2, 2-10 keV, and both bands  There is a possibility that the type-1 AGN sample may contain less luminous AGN than the type-2 AGN due to the deeper depth of the 0.5-2 keV detected sample which is also more sensitive to unobscured AGN. This results in a bias towards less massive galaxies since less luminous AGN have smaller black holes and reside in less massive galaxies.  To perform a fair comparison between the host galaxy stellar mass of type-1 and type-2 AGN, the redshift and luminosity distributions need to be matched. We performed a luminosity and redshift matching process similar to \citet{2019ApJ...878...11Z} and \citet{2023A&A...673A..67G}. Firstly, the sample was divided into redshift and luminosity bins. The redshift was divided into bins of 0.2 between 0.8-1.8 while $\log \lambda L_{\rm 6 \mu m}$ luminosity was divided in bins of 0.25 between $\log L_{\rm 6 \mu m}=43-46$ where each bin contains $N_{L,z}^{\rm t1}$ type-1 AGN and $N_{L,z}^{\rm t2}$ type-2 AGN. The luminosity matching was also performed using the 2-10 keV luminosity. In this case, the 2-10 keV luminosity was divided between \loglx{=42.5-45.5} in bins of 0.25 dex. Secondly, the minimum number of AGN between type-1 and type-2 was chosen $N_{L,z}^{\rm match}=\min(N_{L,z}^{\rm t1}, N_{L,z}^{\rm t2})$ in each $(L,z)$ bin.  Lastly, 10,000 mock samples of type-1 and type-2 AGN were constructed. For each realisation, AGN selected previously can be randomly selected again, following a boot-strap style analysis.

After constructing the mock samples, we perform an Anderson-Darling k-sample test \citep{10.1214/aoms/1177729437} between the mock type-1 AGN sample and the mock type-2 AGN sample. The Anderson-Darling test has the advantage of having increased sensitivity near the edges of the distribution than the Kolmogorov–Smirnov test, where the cumulative distribution asymptotically approaches 0 and 1.\footnote{https://asaip.psu.edu/articles/beware-the-kolmogorov-smirnov-test/} To take into account the stellar mass uncertainties, the stellar mass of each AGN was drawn from a Gaussian distribution with a width equal to the $1\sigma$ uncertainty of the stellar mass.   Here, we used the scipy implementation of the Anderson-Darling k-sample test\footnote{\url{https://docs.scipy.org/doc/scipy/reference/generated/scipy.stats.anderson_ksamp.html}}, which calculates the standardized Anderson-Darling test statistics as well as the p-value. The critical value of the standardised Anderson-Darling statistics that rejects the hypothesis that the stellar mass (or star-formation rates) distribution of type-1 and type-2 AGN at the same at a significance of 0.05 is 1.96. We presented the median Anderson Darling statistics with uncertainties calculated from the 16th and 84th percentile of the distribution. For the p-value, we present the median of the p-value distribution but not uncertainties. This is because the current scipy implementation caps the p-values at 0.25 and 0.001 since the coefficients required to convert the test statistics to p-values are unavailable. If the median p-value is affected by the maximum or minimum p-value, we report the p-value as an upper or lower limit. This cap does not affect the value of the Anderson-Darling statistics. 

We also performed the statistical test on samples limited to 2-10 keV band detection and those with detection in both 0.5-2 and 2-10 keV bands.  Table \ref{tab:adsum} shows the standardized Anderson-Darling test statistics for each exercise.   In all cases, type-2 AGN resides in more massive galaxies than type-1 AGN by $\sim 0.2-0.3$ dex at a 95\% confidence. Our results follow the same trend as recent studies such as \citet{2019ApJ...878...11Z,2021A&A...653A..70M,2022A&A...658A..35K} and \citet{2023A&A...673A..67G} which suggest that type-2 AGN reside in more massive galaxies than type-1 AGN. However,  is in tension with \citet{2019ApJ...872..168S} which concluded that type-1 AGN reside in more massive galaxies, as well as \citet{2012MNRAS.427.3103B} who find no significant difference between obscured and unobscured AGN host.
 
\begin{table}
    \centering
    \caption{Anderson-Darling standardized test statistics for stellar mass.}
    \label{tab:adsum}
    \begin{tabular}{l|c|c|c}
    \hline
    \multicolumn{4}{c}{Stellar Mass Matched with 6\um Luminosity} \\ 
    \hline
    Sample  & Nmatched  & AD-Statistics & P-value\\
    \hline
    All AGN  & 174 & $20.0_{-5.5}^{+6.2}$ & <0.001\\
    2-10 keV & 126 & $17.7_{-5.0}^{5.8}$ & <0.001 \\
    0.5-2 \& 2-10 keV  & 108 & $15.3_{-4.3}^{+5.2}$ & <0.001 \\
    \hline
    \multicolumn{4}{c}{Stellar Mass Matched with 2-10 keV Luminosity} \\ 
    \hline
    Sample  & Nmatched  & AD-Statistics & P-value\\
    \hline
     All AGN & 175 & $18.4_{-5.5}^{+6.7}$ & <0.001 \\
     2-10 keV  & 135 & $13.8_{-4.8}^{5.5}$ & <0.001 \\
     0.5-2 \& 2-10 keV & 107 & $10.6_{-4.0}^{4.8}$ & <0.001\\ 
     \hline
    \end{tabular}
\end{table}

\subsection{The Black Hole mass of Type 1 \& 2 AGN} \label{sec:res:mbh}

In this section, we discuss the black hole mass of  type-1 and type-2 AGN. Single epoch black hole mass estimations using the virial mass relationships are only applicable to broad-line AGN. For type-2 AGN which shows no broad emission lines in the optical/UV spectrum, there is no direct method to estimate the SMBH mass at this redshift. However, the black hole mass can be inferred from the host galaxy stellar mass assuming an $M_{\rm BH}-M_{\rm stellar}$ relationship. 

\begin{figure}
    \centering
    \includegraphics[width=\linewidth]{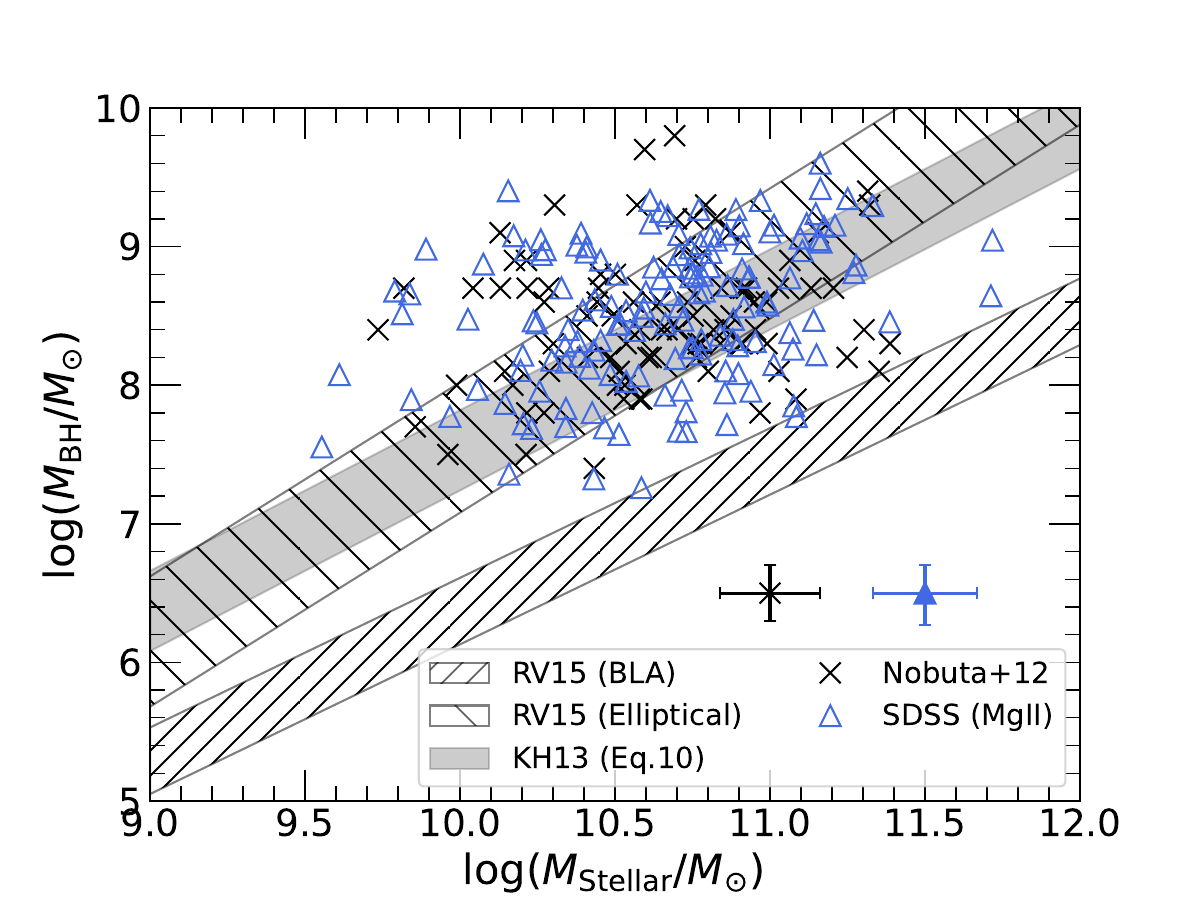}
    \caption{Comparison between the host galaxy stellar mass of type-1 AGN. Black crosses and blue open triangles show measurements obtained from \citet{2012ApJ...761..143N} and spectral analysis, respectively. The data points with error bars show the typical uncertainties of the measurements. The right-hatched and left-hatched areas show the $M_{\rm stellar}-M_{\rm BH}$ relationship for broad-line AGN and early-type galaxies in the early universe from \citet{2015ApJ...813...82R}. The gray shaded area shows the  $M_{\rm stellar}-M_{\rm BH}$ relationship for local early-type galaxies from \citet{2013ARA&A..51..511K}.}
    \label{fig:mstellar_mbh_t1}
\end{figure}

First, we compared the stellar mass and black hole mass of type-1 AGN with local $M_{\rm BH}-M_{\rm stellar}$ from \citet{2013ARA&A..51..511K} and \citet{2015ApJ...813...82R} in Figure \ref{fig:mstellar_mbh_t1}.  The stellar mass to black hole mass ratio for the type-1 AGN is consistent with the $M_{\rm BH}-M_{\rm stellar}$ of local early-type galaxies from both studies but is inconsistent with $M_{\rm BH}-M_{\rm Stellar}$ relationship of broad-line AGN in the local universe.  Our results are similar to previous results which suggest $M_{\rm BH}-M_{\rm stellar}$ of type-1 AGN beyond the local universe is consistent with local passive galaxies\citep{2009ApJ...706L.215J,2013ApJ...767...13S,2020ApJ...888...37D,2021ApJ...909..188S,2023A&A...672A..98M,2020ApJ...889...32S}. This suggest no evolution in the $M_{\rm BH}-M_{\rm stellar}$.  

\begin{figure}
    \centering
    \includegraphics[width=\linewidth]{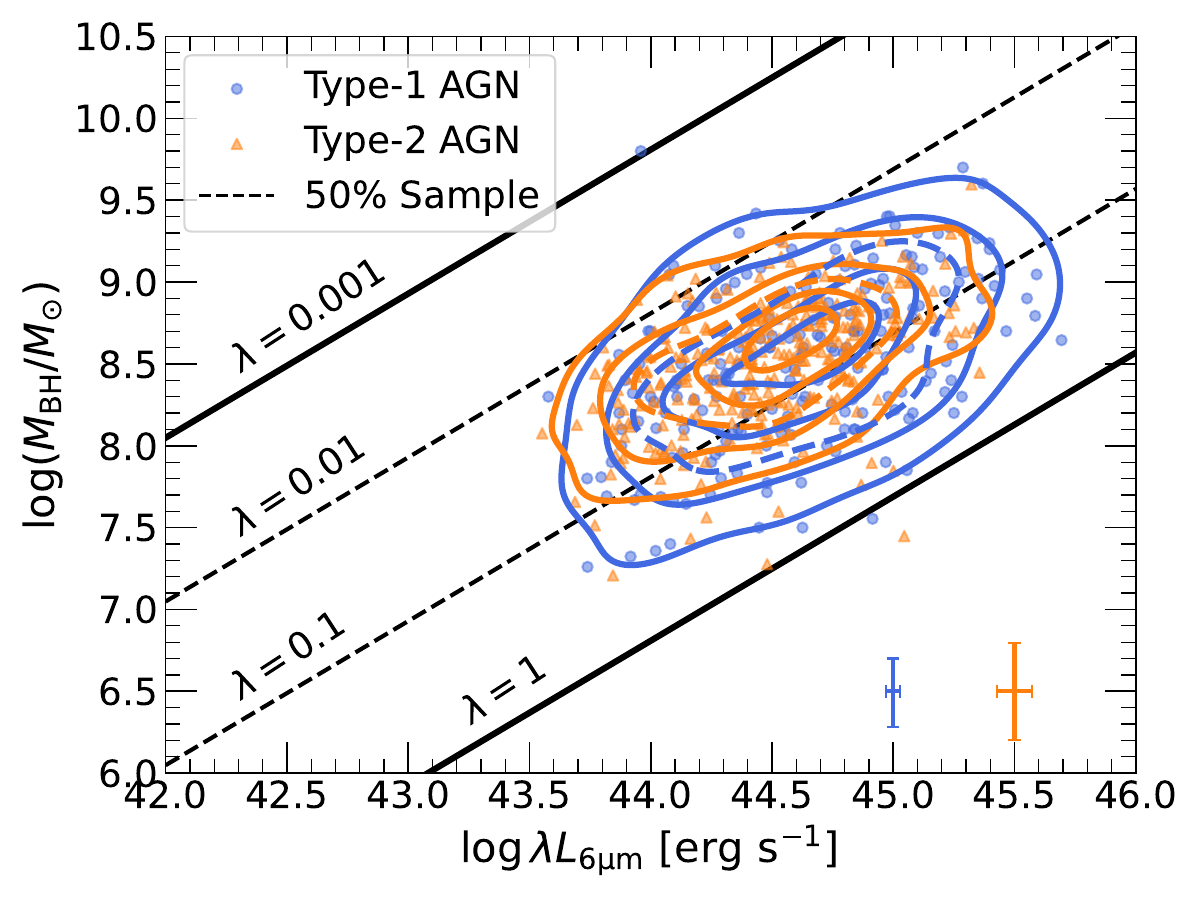}
    \caption{The 6-micron luminosity against black hole mass $M_{\rm BH}$ of type-1 AGN and type-2 AGN shown as blue and orange contours, respectively.  The contour levels show the area which 10-90\% of the datasets lay under. The 50\% level is shown with dashed contours. The lines show the Eddington-ratio assuming the infrared bolometric correction from \citet{2012MNRAS.426.2677R}. The faceless data point with error bars shows the typical uncertainty of the AGN.}
    \label{fig:l6_mbh}
\end{figure}

We used the $M_{\rm BH}-M_{\rm stellar}$ of local early-type galaxies from \citet{2013ARA&A..51..511K} to estimate the black hole mass of type-2 AGN under the assumption that the $M_{\rm BH}-M_{\rm stellar}$ of type-2 AGN is similar to type-1 AGN.  Figure \ref{fig:l6_mbh} shows the distribution in the 6 \um luminosity and black hole mass of type-1 and type-2 AGN. On average, type-2 AGN have an $M_{\rm BH}$ of $\sim 10^{8.5}\ M_\odot$ which is larger than type-1 AGN by $\sim 0.2$ dex. These results may be consistent with AGN in the local universe from the Swift/BAT survey \citep{2022ApJS..261....6K} which estimated the black hole mass of local type-2 AGN using the velocity dispersion relationships. As shown by \citet{2022ApJS..261....9A}  type-2 AGN have more massive black holes than type-1 AGN.

 The choice of the $M_{\rm BH}-M_{\rm stellar}$ relationship can affect the determination of the black hole mass. There is currently no reliable direct method to estimate the black hole mass of type-2 AGN at z=0.8-1.8. Hence, the $M_{\rm BH}-M_{\rm stellar}$ was adopted based on the assumption that type-2 AGN follows a similar $M_{\rm BH}-M_{\rm Stellar}$ relationship as type-1 AGN samples shown in figure \ref{fig:mstellar_mbh_t1}. On one hand, assuming $M_{\rm BH}-M_{\rm Stellar}$ of local broad-line AGN \citep{2015ApJ...813...82R} will decrease the black hole mass of type-2 AGN by approximately 1.5 dex. On the other hand, the $M_{\rm BH}-M_{\rm Stellar}$ of high redshift AGN may have a higher normalization than that of local passive galaxies. \citet{2010ApJ...708..137M} showed that the deviation from the local black hole mass to stellar mass ratio of type-1 evolves with redshift as $(1+z)^{0.68}$. Hence, accounting for the redshift evolution of the $M_{\rm BH}-M_{\rm Stellar}$, the mass of type-2 AGN is larger by $\sim 0.62\log (1+z)\sim 0.27$. 
Recently, there has been the development of a virial mass calibration using the Paschen-$\beta$ emission lines \citep{2017MNRAS.464.1783O} which suggests that type-2 AGN have smaller black hole masses than type-1 AGN by $\sim 0.8-1$ dex \citep{2017MNRAS.468L..97O,2017MNRAS.471L..41R}. While there are differences in the AGN luminosity compared to the sample, future observations of Paschen-$\beta$ emission lines may shed light on the SMBH mass of type-2 AGN. We further discuss how the choice of $M_{\rm BH}-M_{\rm stellar}$ affects the Eddington ratio and our results in section \ref{sec:res:nhledd}.

\subsection{Star Formation Activity} 
\label{sec:res:sfa}

\begin{figure}
    \centering
    \includegraphics[width=\linewidth]{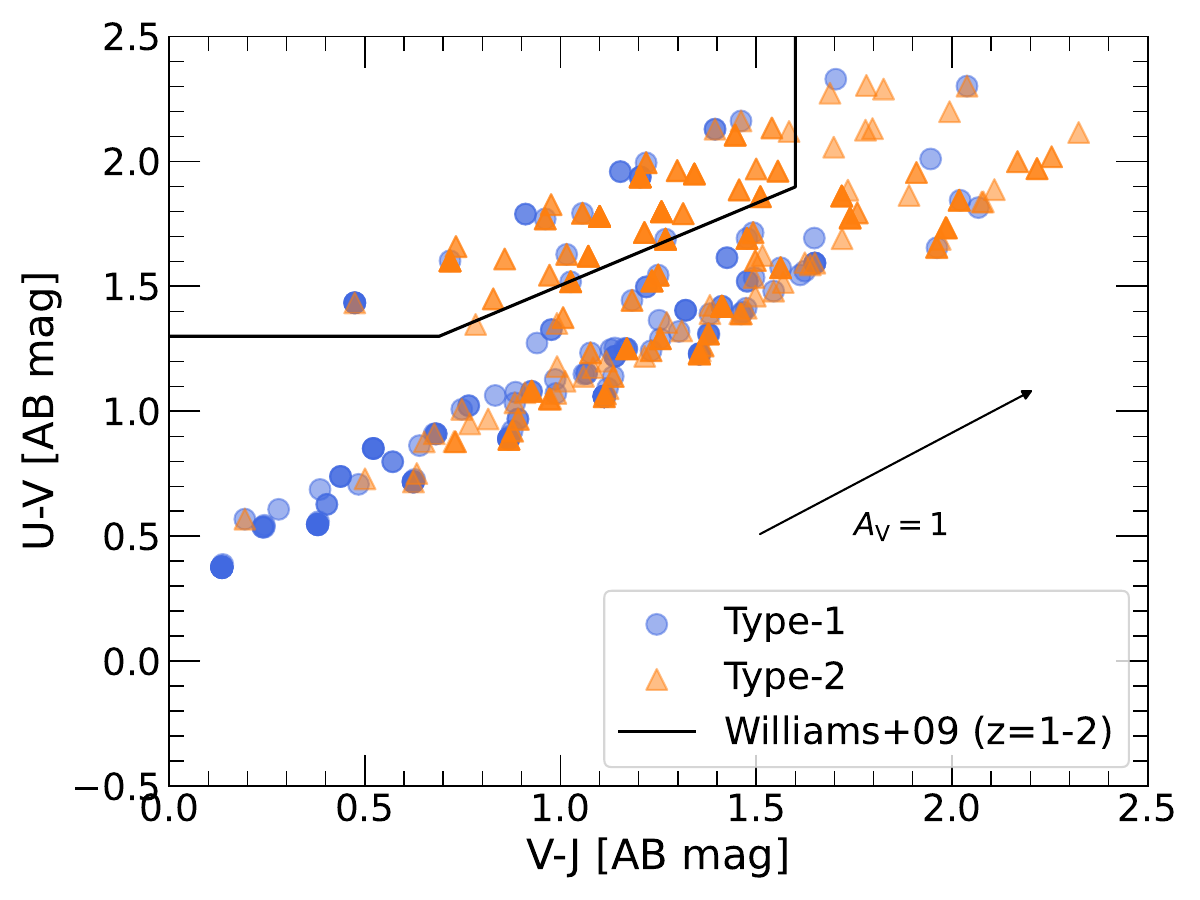}
    \caption{UVJ Diagram of type-1 and type-2 AGN assuming the boundary of \citet{2009ApJ...691.1879W}. The reddening vector with $A_{\rm V}=1$ is shown on the right. }
    \label{fig:uvj}
\end{figure}

In this section, we discuss the star-formation properties of the AGN host galaxies. Firstly, we examined the colours in the restframe UVJ diagram. The UVJ diagram classifies galaxies into two major groups including; star-forming galaxies on the bottom right side and passive galaxies on the top left wedge.  Figure \ref{fig:uvj} shows the colour distribution of the host galaxy of type-1 and type-2 AGN on the UVJ diagram with the AGN component removed. We adopted the star-forming/passive boundary of $z=1-2$ galaxies from \citet{2009ApJ...691.1879W} to classify star-forming and quiescent galaxies. The majority of type-1 and type-2 AGN reside in the star-forming portion of the diagram while 22(11\%) of type-1 AGN and 109(37\%) of type-2 AGN  reside in the passive region but close to the boundary of the diagram. A portion of the type-1 AGN occupies the bluer part of the UVJ diagram at $V-J<0.6$ and $U-V<1$. We note that the overlapping points in the colours of type-1 and type-2 host seen in the passive zone are because both type-1 and type-2 AGN were fitted with the same best-fitted host galaxy model. 

\begin{figure*}
    \centering
    \includegraphics[width=\linewidth]{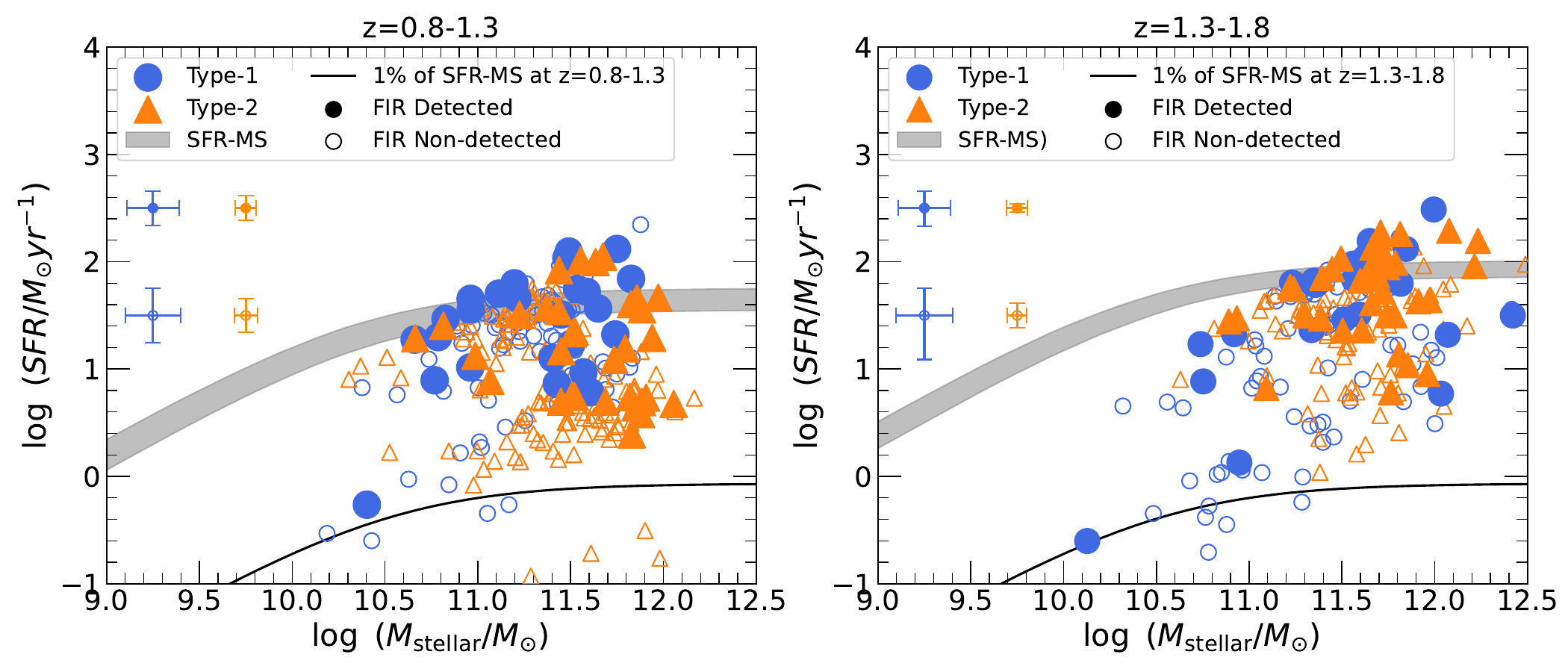}
    \caption{The ${\rm SFR}-M_{\rm stellar}$ distribution at $z=0.8-1.3$ (left) and at $z=1.3-1.8$ (right). FIR-detected AGNs are shown as filled data points while those with FIR-nondetections are shown with open symbols.  The star formation main sequence of \citet{2023MNRAS.519.1526P} is shown as the grey area where the with represents $1\sigma$ the confidence interval of the relationship. }
    \label{fig:sfrms}
\end{figure*}

Figure \ref{fig:sfrms} shows the distribution of type-1 and type-2 AGN in the ${\rm SFR}-M_{\rm stellar}$ diagram. We separated the samples into two redshift bins to mitigate the effect of the cosmological evolution of the star-formation main sequence towards cosmic noon and compare it with the star-formation main sequence of \citet{2023MNRAS.519.1526P}.  To be consistent with the IMF assumed in of \citet{2023MNRAS.519.1526P}, we converted the stellar mass to that with the Kroupa IMF \citep{2001MNRAS.322..231K} with a 1.06 conversion factor \citep{2023MNRAS.519.1526P}.  We also adopted the 1\% of the main-sequence SFR as the fiducial quenched limit (black solid line).

For the majority of the AGN population, the star-formation rates are consistent with the SFR of main-sequence galaxies of the same stellar mass or are lower but not quenched. We performed an Anderson-Darling test to the similarity between the SFR and the specific star formation rate (SSFR) of type-1 and type-2 AGN. We performed the same redshift and luminosity matching as described in section \ref{sec:res:mstellar}, but in addition also matched the host galaxy stellar mass distribution since both SFR and SSFR is a function host galaxy stellar mass \citep{2005ApJ...633L...9F,2011ApJ...730...61K,2015A&A...579A...2I}. We matched the stellar mass distribution between $\log M_{\rm stellar}=9-12$ in bins of 0.5 dex. Both FIR-detected and FIR-non-detected AGN were used in the analysis since not enough FIR-detected AGN were available for statistical analysis after matching redshift, AGN luminosity and stellar mass. 

The results suggest no difference between the SFR and SSFR of type-1 and type-2 AGN at a 95\% confidence and are summarized in table \ref{tab:ssfrtest}. Our results are consistent with \citet{2019ApJ...878...11Z,2021A&A...646A.167M,2021A&A...653A..70M,2022A&A...658A..35K,2019ApJ...872..168S} who concluded that the SFR of type-1 and type-2 AGN are statistically consistent with each other and are consistent with typical main-sequence galaxies or slightly below the main-sequence. 

\begin{table}
    \centering
    \caption{Anderson-Darling standardized test statistics of type-1 and type-2 AGN for the host galaxy SFR and SSFR.}
    \label{tab:ssfrtest}
    \begin{tabular}{l|l|l||l}
         \hline
         \multicolumn{4}{c}{Star formation Rate Matched with 6\um Luminosity} \\ 
         \hline
         Sample   & Nmatch & AD-Statistics & P-value\\
         \hline 
         All &  125 &  $1.1_{-1.1}^{+1.8}$ & 0.12 \\
         2-10 keV & 86 & $2.0_{-1.4}^{+2.1}$ & 0.05 \\
         0.5-2 \& 2-10 keV & 69 & $0.8_{-1.0}^{+1.6}$ & 0.16 \\
         \hline
         \multicolumn{4}{c}{Star formation Rate Matched with 2-10 keV Luminosity} \\ 
         \hline
         Sample   & Nmatch & AD-Statistics & P-value\\
         \hline
         All  & 131 &  $1.3_{-1.2}^{+2.1}$ & 0.10 \\
         2-10 keV   &  95 & $0.4_{-0.7}^{+1.4}$ & 0.23 \\
         0.5-2 \& 2-10 keV  & 73 & $0.2_{-0.6}^{+1.1}$ & >0.25\\
         \hline
         \multicolumn{4}{c}{Specific Star Formation Rate Matched with 6\um Luminosity} \\ 
         \hline
         Sample  & Nmatch & AD-Statistics & P-value \\
         \hline
         All  & 125 & $1.0_{-0.9}^{+1.5}$ & 0.13 \\
         2-10 kev & 86 & $1.6_{-1.2}^{+1.9}$ & 0.07\\
         0.5-2 \& 2-10 keV & 69 & $1.6_{-1.2}^{+2.0}$ & 0.09 \\
         \hline
         \multicolumn{4}{c}{Specific Star Formation Rate Matched with 2-10 keV Luminosity} \\ 
         \hline
         Sample  & Nmatch & AD-Statistics & P-value \\
         \hline
         All   & 131 & $1.2_{-0.9}^{+1.6}$ & 0.10 \\
         2-10 kev  & 95 & $0.7_{-0.7}^{+1.1}$ & 0.17 \\
         0.5-2 \& 2-10 keV  & 73 & $0.8_{-0.8}^{+1.1}$ & 0.16 \\
         \hline
    \end{tabular}
    
\end{table}

\subsection{$N_H-\lambda_{\rm Edd}$ Diagram} \label{sec:res:nhledd}

\begin{figure}
    \centering
    \includegraphics[width=\linewidth]{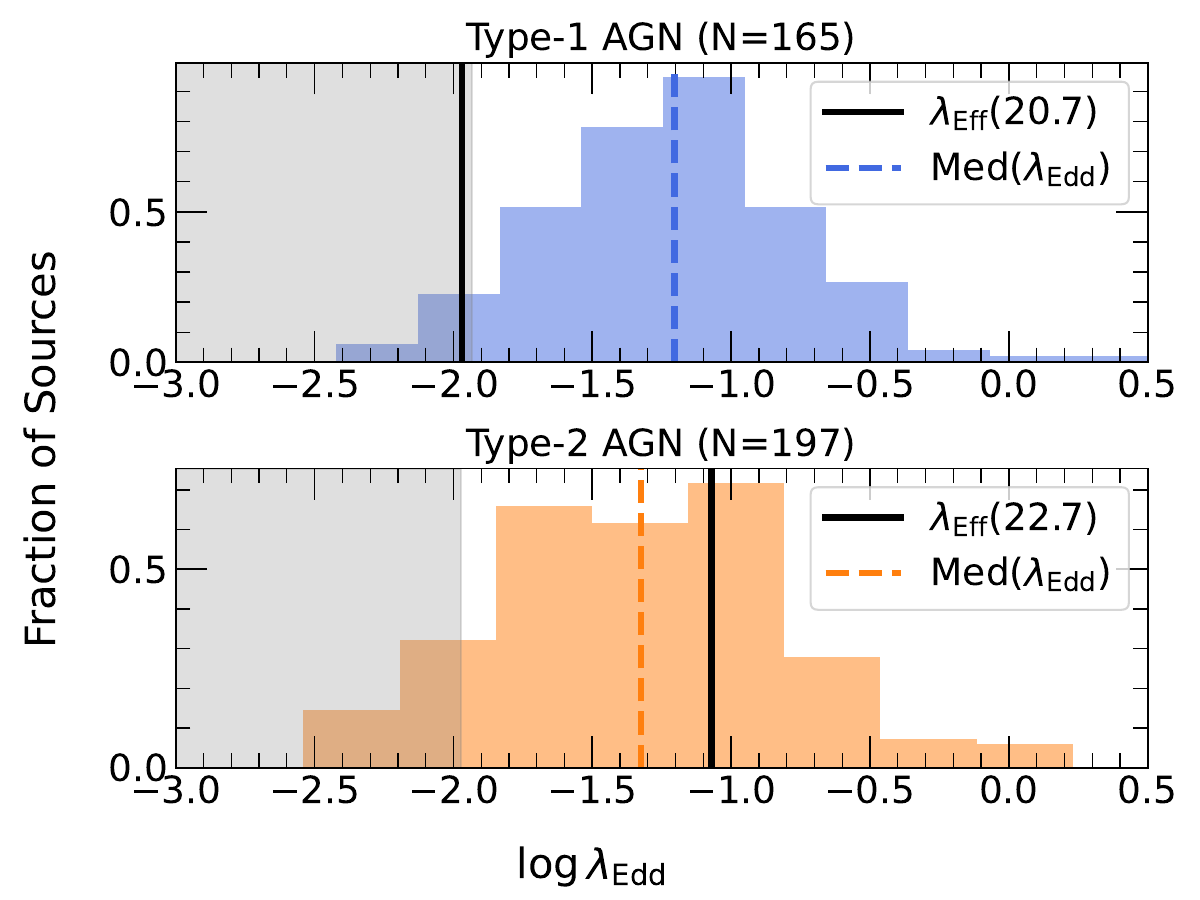}
    \caption{The observed Eddington-ratio distribution of Type-1 AGN (top) and type-2 AGN (bottom). The thick black lines show the effective Eddington ratio at the median hydrogen column density of each sample assuming a dust abundance of 0.3 galactic value. The dashed line shows the median Eddington ratio of each sample. The grey-shaded area shows the range where the flux limit strongly affects the distribution.}
    \label{fig:edd_dist}
\end{figure}

In this section, we present the obscuration properties of type-1 and type-2 AGN. Here, we limited the sample to 362 AGN with detection in the 2-10 keV band including 165 type-1 AGN and 197 type-2 AGN. We examined the observed Eddington ratio distribution for type-1 and type-2 AGN detected in the 2-10 keV band in figure \ref{fig:edd_dist}. We adopted the hard X-ray bolometric correction from \citet{2020A&A...636A..73D} to calculate the bolometric luminosity. The bolometric correction is expressed as $$K_{\rm X}(L_{\rm bol})=a\bigg[1 + \bigg(\frac{(\log L_{\rm Bol}/L_\odot)}{b}\bigg)^c \bigg]$$ where a, b, and c are constants of the bolometric correction described in table \ref{tab:bolcorr}, $K_{\rm X}(L_{\rm bol})$ is the hard X-ray bolometric correction, and $L_{\rm Bol}$ is the bolometric luminosity. The uncertainty of the k-correction was calculated by the propagation of the parameter uncertainties, including the scatter in the relationship.

\begin{table}
    \caption{Bolometric correction constants from \citet{2020A&A...636A..73D}.}
    \centering
    \begin{tabular}{ccccc}
        \hline
        AGN & a & b & c & scatter\\
        \hline
        Type-1 & $12.76 \pm 0.13$ & $12.15\pm0.01$ & $18.78 \pm 0.14$ &  0.26 \\
        Type-2 & $10.85 \pm 0.08$ & $11.90\pm0.01$ & $19.93 \pm 0.29$ &  0.27 \\
        \hline
    \end{tabular}
    \label{tab:bolcorr}
\end{table}

The effective Eddington ratio for the median hydrogen column density is shown with the thick black line. The majority of type-1 AGN are above the effective Eddington ratio while the opposite is true for type-2 AGN. In addition, the Eddington ratio distribution of type-2 AGN shows a skewed distribution towards a lower Eddington ratio while the type-1 AGN shows a symmetric distribution.  There is no significant difference between the median Eddington ratio between type-1 and type-2 AGN. For both samples, the typical completeness limit is at an Eddington ratio of $\log \lambda_{\rm  Edd}\sim -1.9$. 

\begin{figure*}
    \centering
    \includegraphics[width=\linewidth]{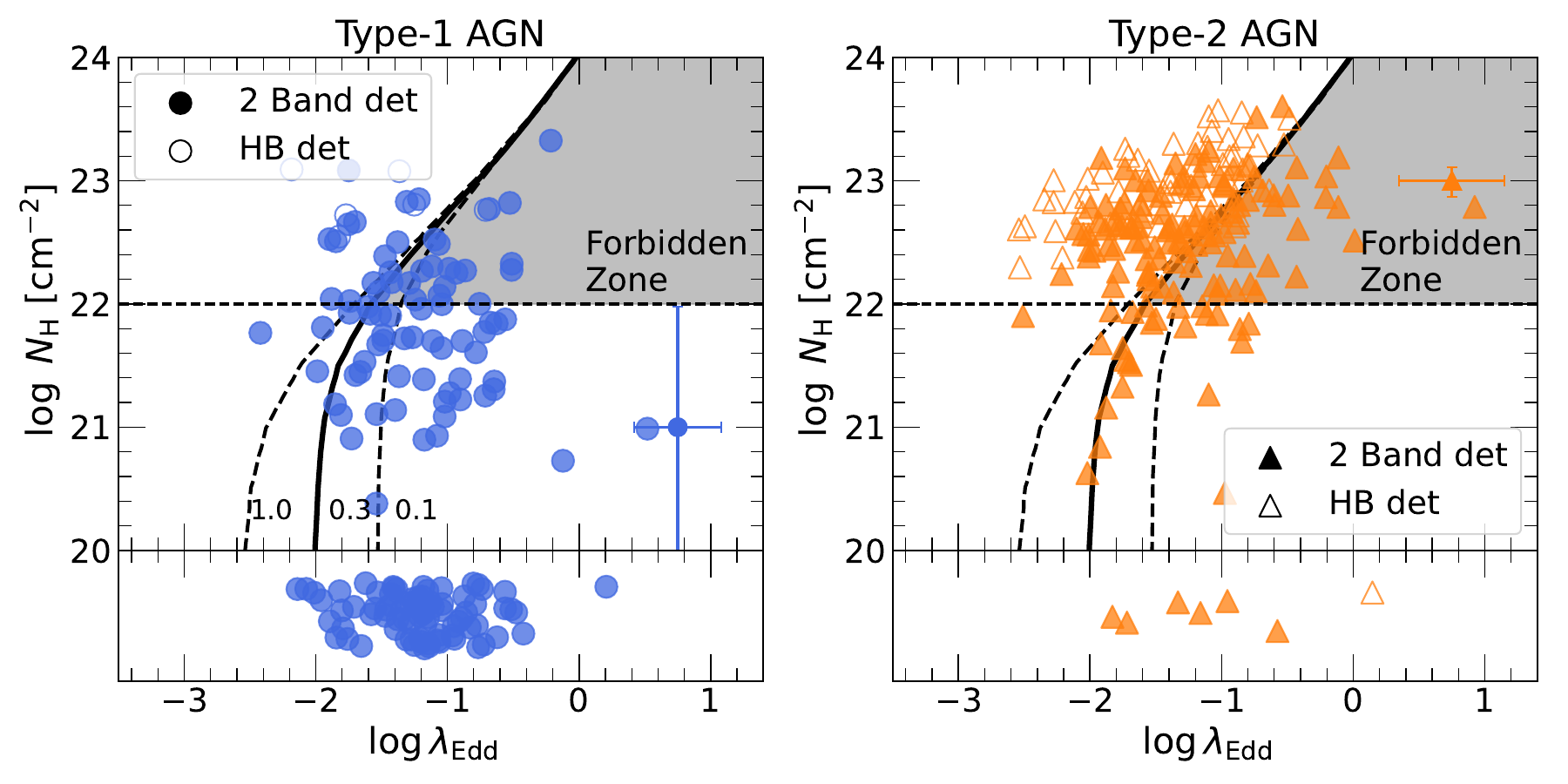}
    \caption{The $N_H-\lambda_{\rm Edd}$ Diagram of type-1 AGN (left, blue round symbols) and type-2 AGN (right, orange triangles). We show the hydrogen column density uncertainty assuming an intrinsic column density of \lognh{=21} and \lognh{=23} and the typical Eddington-ratio uncertainty for type-1 and type-2 AGN with the data points with errorbars. In both panels, the curved thick and dashed line shows the effective Eddington ratio assuming a dust abundance of 1, 0.3, and 0.1 galactic value, from left to right. The boundary of the forbidden zone is defined assuming the dust abundance of 0.3 galactic value. The horizontal dashed line shows the fiducial ISM obscuration level in the local universe. The grey area shows the forbidden zone. The bottom panel shows the Eddington-ratio distribution of AGN which the Hydrogen column density can not be constrained with the X-ray hardness ratio.}
    \label{fig:nh_lambda}
\end{figure*}

Figure \ref{fig:nh_lambda} shows the distribution of type-1 and type-2 AGN at redshift 0.8-1.8 in $\log N_{\rm H}-\lambda_{\rm Edd}$ diagram on the left and right-hand side, respectively. The bottom panel of each plot shows the distribution of sources whose hydrogen column density cannot be constrained with the X-ray hardness ratio because they have a softer hardness ratio than the intrinsic power-law model ($\Gamma=1.8$). Their position is shown along the Eddington ratio axis but randomly distributed vertically for clarity. We adopted the effective Eddington ratio lines from \citet{2008MNRAS.385L..43F} for simplicity. \citet{2022MNRAS.517.5069A} showed that the boundary of the effective Eddington ratio depends mainly on the dust grain abundance and dust size. This strongly affects the boundary below \lognh{\leq22} but not higher column densities unless the dust abundance is significantly lower than 0.1 solar abundance. In Figure \ref{fig:nh_lambda}, we show the dependence of the effective Eddington limits on the dust abundance of 0.1 to 1 galactic dust abundance. In this work, we define the forbidden zone boundary based on the line assuming a dust abundance of 0.3 galactic value.

For type-1 AGN, only 78 of the 165 2-10 keV detected type-1 AGN  have $\log N_{\rm H}$ which can be constrained with the hardness ratio. However, since broad lines are observed among this sample, the hydrogen column density for the majority should be lower than \lognh{\sim22}, otherwise the dust attenuation would extinguish AGN emission. Some type-1 AGN has hydrogen column density larger than \lognh{=22} but since the optical spectra show broad lines, it implies that the X-ray absorption is caused by material with low dust content or free of dust.  We discuss this further in \ref{sec:disc:t1obs}

Most type-2 AGNs have a hydrogen column density between \lognh{\geq 22-24} consistent with the typical classification of X-ray obscured AGN. Among the type-2 AGN with \lognh{=22-24}, 25\% reside in the forbidden zone. At face value, it suggests that these type-2 AGNs are actively producing dusty outflows. An alternative explanation is that the X-ray emission is affected by heavy ISM absorption associated with kpc scale gas \citep{2015MNRAS.451...93I}. We discuss this further in section \ref{sec:disc:t2obs}.

As mentioned in section \ref{sec:res:mbh}, the choice of the $M_{\rm BH}-M_{\rm stellar}$ relationship can affect the black hole mass of type-2 AGN, and therefore, also the Eddington ratio. Assuming the $M_{\rm BH}-M_{\rm stellar}$ relationship of local broad-line AGN \citep{2015ApJ...813...82R} will increase the Eddington ratio by a 1.5 dex. In such a case, most of the type-2 AGN  will appear in the forbidden region near the Eddington limit above the effective Eddington ratio limit. On the other hand, the redshift depended $M_{\rm BH}-M_{\rm stellar}$ presented by \citet{2010ApJ...708..137M} will reduce the Eddington ratio for type-2 AGN by 0.27 dex.  In this case, the fraction of AGN in the forbidden zone is reduced.

\subsection{The Colors of Type-1 and Type-2 AGN} \label{sec:res:agncolor}
 In this section, we investigate the effects of ISM dust and gas within the line of sight on the restframe colours of type-1 AGN using a toy model. In principle, the line of sight of type-1 AGN passes through both the host galaxy ISM and the polar dust region.  Therefore, the X-ray absorption can be due to the ISM with modest column density in the line of sight. Since the hydrogen column density of the majority of type-1 AGN can not be constrained, we examined the effect of ISM obscuration on the restframe colours of the AGN. 

We constructed a toy model to simulate the observed colours when the AGN optical emission of a type-1 AGN is affected by ISM attenuation as a function of ISM hydrogen column density. In CIGALE, the polar dust attenuation in the AGN model is applied to the AGN component and independent from the dust attenuation applied to the stellar component. Since distinguishing between polar dust and the ISM is not straightforward, we do not distinguish between polar dust and the ISM in the line-of-sight towards the AGN. We used this polar dust extinction to represent the attenuation by the kpc-scale ISM ($<1$ kpc) or circumnuclear scale ($\sim 100$ pc) in the case of \lognh{\sim 24}. We adopted the SMC extinction curve for the dust attenuation applied to the AGN component and converted $\log N_{\rm H}$ to colour excess ($E(B-V)$) assuming $A_{\rm V}/N_{\rm H}$ from \citet{2009AJ....137.5099D} and $R_{\rm V}=2.93$ from \citep{1992ApJ...395..130P}. This setup represents a highly simplified 2-step distribution of dust attenuation in the host galaxy, where the centre part of galaxies have larger optical attenuation in the inner regions \citep{2019ApJ...887..204J,2020MNRAS.495.2305G} as well as larger gas surface density \citep{2012ApJ...756..183B}. 

We adopted two sets of host galaxy models from the best-fitted parameters from the SED fitting of type-1 AGN. Firstly, a 3 Gyr old stellar population with e-folding time of 0.5 Gyr and secondly, a 1.5 Gyr old stellar population with e-folding time of 6 Gyr. We used the dust extinction curves of \citet{1989ApJ...345..245C} with $R_{\rm V}=3.1$ and set the colour excess of 0.25, 0.6, and 1\footnote{E\_BV\_lines parameter in the dustatt\_modified\_starburst module}. The choice of colour excess shown here is based on the 16th, 50th, and 84th percentile colour excess of type-2 AGN in our sample. This corresponds to a V-band attenuation of 0.77, 1.86, and 3.1 magnitudes, respectively. To realistically set the ratio between the host galaxy luminosity and the AGN luminosity, we calculated the AGN fraction in 2-10 \um of the best-fitted type-1 AGN model. We set the AGN fraction in the models accordingly. 

\begin{figure}
    \centering
    \includegraphics[width=\columnwidth]{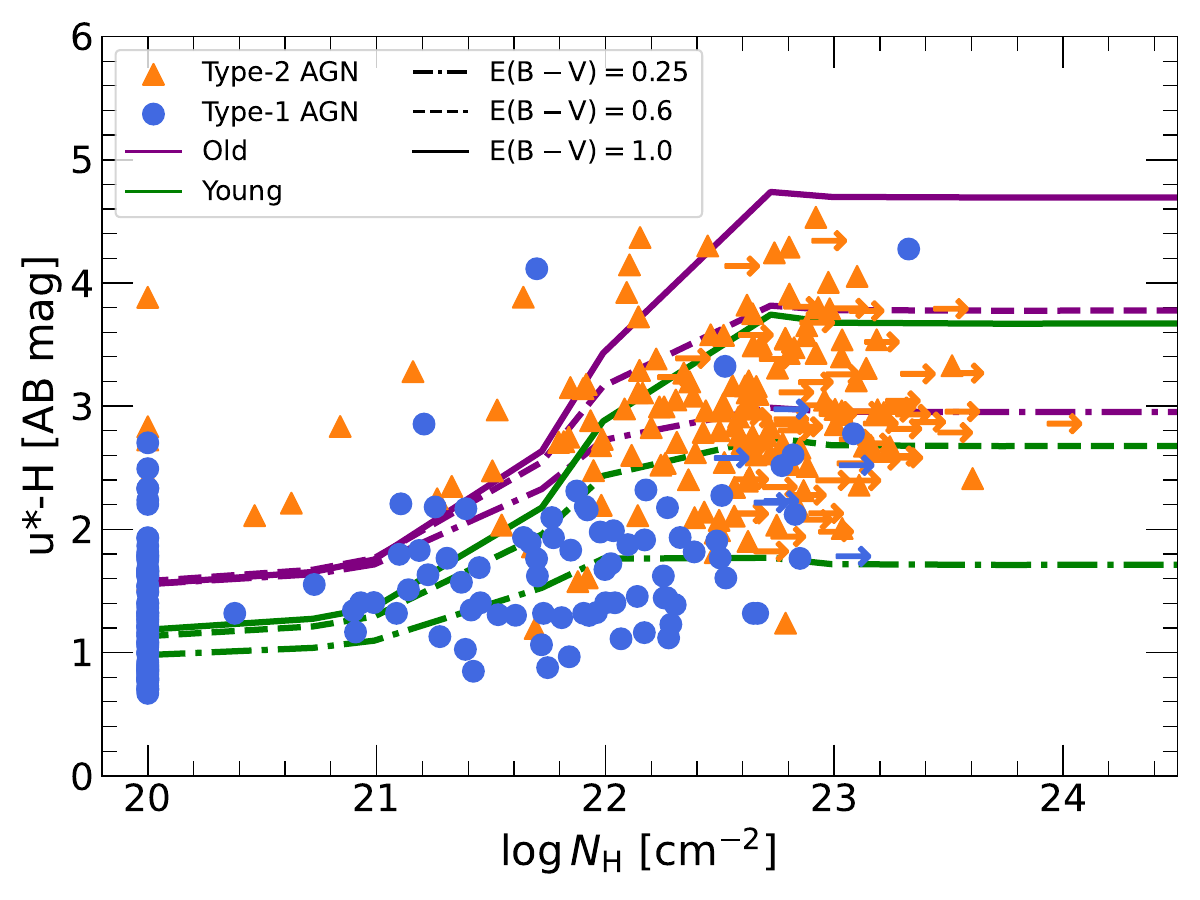}
    \caption{Hydrogen column density against rest-frame u*-H colour of type-1 and type-2 AGN. shown as blue round and orange triangles, respectively. Model colours of AGN host galaxy of an old (3 Gyr stellar age with 0.5 Gyr e-folding time) and young (1.5 Gyr stellar age with 6 Gyr e-folding time) are shown in purple and green lines, respectively. The dot-dashed, dashed, and thick lines show the models with colour excess of $\rm E(B-V)=0.25,0.6$, and 1, respectively }
    \label{fig:simcolor}
\end{figure}

In figure \ref{fig:simcolor}, we present the $u^*-H$ colour distribution of type-1 AGN against the total hydrogen column density. We plot the $u^*-H$ colours of the two models as a function of kpc-scale ISM-only hydrogen column density. The colours of type-2 AGN are shown as a reference for host galaxy-dominated colours. With increasing kpc-scale ISM obscuration which affects the AGN emission,  the contamination by the host galaxy increases as seen as the increasing red colours at \lognh{\sim 21.5-22}.  Above \lognh{=22}, the colours become dominated by the host stellar emission as there is no longer any change in the colours.

For type-1 AGN with \lognh{<22}, the colours are consistent with those which are dominated by the blue unobscured AGN emission. Some type-1 AGN have hydrogen column density larger than \lognh{>22} but have colours consistent with the unobscured type-1 AGN or obscured AGN in host galaxies with young stellar populations with low dust attenuation. However, the latter case cannot explain the broad emission lines detected in the optical spectra of the type-1 AGN since the latter case suggests the AGN emission is extinct due to the large kpc-scale dust attenuation. This implies that the line of sight towards the type-1 AGN is gas-rich but dust-free.

\section{Discussions}

\subsection{The Obscuration of Type-1 AGN}
\label{sec:disc:t1obs}

In this section, we discuss the X-ray absorption of type-1 AGN on the $N_{\rm H}-\lambda_{\rm Edd}$ diagram. As previously mentioned, the detection of broad emission lines in the optical spectra implies that the obscuration in the line of sight is modest compared to type-2 AGN. In section \ref{sec:res:agncolor}, we showed that the restframe colours of type-1 AGN with \lognh{<22} can be explained with mild ISM extinction while type-1 AGN with X-ray absorption \lognh{>22} can be explained if they occur from dust-free gas. Dust-free obscuration has been discussed in the literature under two major scenarios. 

The first is due to lines of sight that intersect with dust-free gas clouds within the dust sublimation radius or near the edge of the torus. \citet{2014MNRAS.437.3550M} showed that obscured broad-line AGN makeup $\sim 10-20\%$ of the luminous AGN population (\loglx{>44}). \citet{2019ApJ...870...31I} investigated the X-ray and mid-infrared dust covering factor of Swift/BAT AGN. They found that for typical AGN with $\log L_{\rm bol}\rm \ [erg s^{-1}]>42.5$ the covering factor of X-ray absorbing material is larger than that of the mid-infrared emitting dusty material.  \citet{2018MNRAS.479.5022L} concluded that partial obscuration in lines-of-sight near the torus edge can result in dust-free obscuration. Another possibility is that dust-free lines of sight are due to a gas-rich extended time variable dust sublimation zone within the nuclear torus \citep{2023ApJ...950...72K}.

The second scenario is due to accretion disk winds within the dust sublimation radius which coincides with the line-of-sight. Since these winds occur within the dust sublimation radius, their content is dust-free ionised gas. An example of such a case is broad-absorption line quasars (BAL-QSO) which tend to show complex ionised X-ray absorption and broad absorption lines associated with CIV, MgII, or SiV \citep{2002ApJ...567...37G,2006ApJ...644..709G}. We do not discuss the statistical significance among the sample since most of the sample is at redshift below that which CIV can be detected in the optical spectrum but  5 type-1 AGN show absorption associated with the CIV emission line. One of them has a hydrogen column density of \lognh{= 22.5} assuming a neutral photoelectric absorber (phabs) while the hydrogen column density of the remaining AGN cannot be constrained. BAL-QSO tend to have more complex obscuration including both ionised and partial obscuration \citep{2006ApJ...644..709G} which requires X-ray spectral analysis with a higher S/N ratio.

\subsection{The Obscuration of Type-2 AGN}
\label{sec:disc:t2obs}

In this section, we discuss the X-ray absorption of type-2 AGN on the $N_{\rm H}-\lambda_{\rm Edd}$ diagram. Unlike type-1 AGN, the lack of broad emission lines and the host galaxy-dominated continuum require large amounts of dust along the line of sight. In figure \ref{fig:edd_dist}, the Eddington ratio of type-2 AGN is skewed towards lower Eddington-ratios than type-1 AGN. Among all the 2-10 keV detected type-2 AGN with \lognh{>22}, only 25\% have an Eddington ratio above the critical Eddington ratio limit assuming a gas-to-dust ratio of 0.3 galactic value. This indicates that the majority of obscuration among type-2 AGN in this redshift range is due to a radiation-pressure regulated dusty torus as seen in AGN in the local universe \citep{2022ApJS..261....9A,2022ApJ...938...67R}.  Our results are also consistent with evolutionary models of the obscured fraction of AGN which include the contribution to the obscured fraction by the host galaxy ISM. As shown in \citet{2022A&A...666A..17G} and \citet{2023MNRAS.tmp.3140A}, below redshift 2-3, the AGN dusty torus is still an important component which contributes to the Compton-thin obscured fraction of AGN.

In figure \ref{fig:nh_lambda}, some of the type-2 AGN have an Eddington ratio above the critical Eddington limit of dusty gas which may imply that they are actively producing outflows. \citet{2020MNRAS.495.2652L} suggest that the time scale in which dusty outflows clear out Compton-thick line of sight (\lognh{=24}) is very short ($\sim 10^4$ yr) compared to the lifetime of the AGN ($\sim 10^{7-8}$ yr) and for lower hydrogen column density, the time scale should be shorter. \citet{2021ApJ...906...21J} estimated that $\leq 20\%$ of the AGN population should occupy the blow-out region of the $N_{\rm H}-\lambda_{\rm Edd}$ diagram. Hence, only a small fraction of the entire sample of type-2 AGN should occupy the forbidden zone. The large number of AGN in the forbidden zone can be explained if the X-ray absorption in these AGN is associated with the heavy ISM absorption. Therefore, the obscuration can be kept at accretion rates above the effective Eddington ratio.

We attempted to quantify the increasing incidence of ISM-obscured AGN due to the cosmological evolution of the gas fraction by comparing the fraction of type-2 AGN within the forbidden zone with those in the local universe. For our local AGN sample, we used the second data release of the  Swift-BAT AGN Spectroscopic Survey (BASS DR2, \citet{2022ApJS..261....2K,2022ApJS..261....4O}), which is a highly complete sample of AGN in the local universe ($z<0.1$) thanks to detection of hard X-ray ($>10$ keV) and high spectroscopic completeness thanks to dedicated follow up efforts. We limited the sample to 681 objects with non-beamed and non-lensed AGN that have black hole mass determined from broad emission lines or stellar velocity dispersion. We added the hydrogen column density from \citet{2017ApJS..233...17R} and limited the sample to those that have similar redshift ($|\Delta z|<0.01$). We adopted the spectral classification presented in \citet{2022ApJS..261....2K} to classify the AGN into type-1 or type-2.

In the local universe, the fraction of Compton-thin type-2 AGN in the forbidden zone among AGN with Eddington ratio larger than $\lambda_{\rm Edd}\geq -2$ is $11\pm3 \%$. If we consider Seyfert 1.9  as an ISM reddened type-1 AGN based on the detection of broad $\rm H\alpha$ emission line then the fraction is reduced to $7\pm2\%$. In comparison, the fraction of type-2 AGN at z=0.8-1.8 in the forbidden zone with Eddington ratio larger than $\lambda_{\rm Edd}\geq -2$ is $28\pm5\%$.  
Our results are consistent with \citet{2022A&A...661A..15T} who also found an increased fraction of AGN in the forbidden zone with redshift. This is overall consistent with an increase in ISM obscuration due to the cosmological evolution of the galaxy's ISM.

Since ISM gas is dusty, large ISM hydrogen column density also affects the AGN optical/UV emission \citep[e.g.][]{2021A&A...650A..75G}. As a brief reminder, our sample of type-2 AGN was selected based on their extended morphology which implies that the AGN does not outshine the host. Because the Eddington ratio of type-2 AGN in the forbidden zone is above the effective Eddington ratio, the dusty torus obscuration cannot be maintained. However, the AGN can appear as a type-2 AGN since the heavy ISM dust attenuation reduces the contrast between the host and the AGN light. For hydrogen column densities of \lognh{\sim 22.5}, the V-band attenuation ranges from $A_{\rm V}\sim4.7-15$ mag depending on the assumption of gas-to-attenuation ratios \citep{2009AJ....137.5099D,2017MNRAS.471.3494Z} and the UV attenuation will be 2-5 times larger depending on ISM extinction curves \citep{1989ApJ...345..245C,2000ApJ...533..682C}. 

This is similar to the local intermediate Seyfert galaxies (e.g. Seyfert 1.5-1.8s, \citet{1995ApJ...454...95M}) but with larger hydrogen column density and with larger observed frequency. In the most severe case, the ISM obscuration may be enough to attenuate a large fraction of the stellar emission as seen in heavily obscured submillimeter galaxies (SMG). Among our sample of AGN, we found 16 of the type-2 AGN were not detected in the HSC bands but were detected in the NIR bands. Their column density exceeds \lognh{=22} and the stellar component $E(B-V)\sim 1$, which corresponds to a V-band attenuation of 3.1 magnitudes.  However, the V-band attenuation $A_{\rm V}$ within the kpc-scale could be larger \citep{2019ApJ...887..204J,2020MNRAS.495.2305G} than the V-band attenuation from SED fitting which represents the average optical attenuation over the whole galaxy. Such large amounts of dust attenuation may suggest that these AGN reside in host galaxies with large dust content and similar to SMG-AGN at $z\geq2$ which show heavy absorption from the ISM components \citep{2019A&A...623A.172C,2020A&A...636A..37D,2020MNRAS.494.3828D}.

\subsection{Star-formation Activity and Obscuration}

In this section, we discuss the properties of the star formation of type-1 and type-2 AGN in the context of the evolutionary scenario. Intense star formation is expected to occur simultaneously with obscured AGN activity following a gas-rich merger event as seen in local ULIRGs. If obscuration triggered by a gas-rich merging event is common in the AGN population at high redshift, the SFR in type-2 AGN should be more intense than unobscured AGN \citep{2006ApJS..163....1H,2008ApJS..175..356H}. 

However, firstly the statistical analysis suggests no significant difference between the SFR or SSFR of type-1 and type-2 AGN. Secondly, the SFR of type-1 and type-2 AGN are consistent with the SFR of galaxies in the star formation main sequence.  We conclude that the majority of type-2 AGN are obscured by a radiation pressure-regulated torus or obscuration due to the cosmological evolution of gas in galaxies and not kpc-scale obscuration as a result of gas-rich mergers. 

Although our results support the scenario of the regulated dusty torus, selection effects may affect our result. During the coalescent phase which corresponds to the peak of the accretion and SFR in the major merging event, the large influx of gas towards the nuclear regions may result in near Compton-thick absorption \citep{2018MNRAS.478.3056B}, with which they can be missed from X-ray selection due to their faintness.  As discussed in \citet{2022MNRAS.517.2577A} and \citet{2015ApJ...802...50C}, infrared-selected obscured quasars have higher star formation than unobscured quasars. Recently, \citet{2024MNRAS.527L.144A} showed that a large fraction of submillimeter selected AGN with high star formation rates are heavily obscured by a compact ISM component, indicating that they are buried in a compact starbursting region.

From the perspective of hydrodynamical simulations of galaxy mergers, \citet{2022ApJ...936..118Y} showed that the star formation rate, colours, and hydrogen column density of simulated galaxy mergers can show significant variability along the merger sequence which can result in a large range scatter in SFR. If we are truly missing heavily obscured AGN during the peak of the SFR and obscuration due to the heavy photoelectric absorption, our sample of type-1 and type-2 AGN may represent AGN in the early stage of coalescence where the AGN is triggered by gravitational disturbance of nearby galaxies and the absorption is due to torus or nuclear ISM obscuration and the SFR and SSFR are consistent with the main-sequence galaxies. For type-1 AGN, there is also the possibility of the AGN being in the post-blowout stage after a decline in SFR but not quenched due to the longer time scale of star formation compared to the AGN. High angular resolution deep imaging data and spectroscopic follow-up of a combined X-ray and infrared sample are needed to understand further the prevalence of large-scale merger-induced ISM obscuration among the AGN population.

\section{Conclusions}

In this study, we used the deep multiwavelength photometric and spectroscopic data in the HSC-Deep XMM-LSS survey field to investigate the incidence of obscuration, stellar mass, and SFR among X-ray selected type-1 and type-2 AGN at $z=0.8-1.8$. To determine the Eddington ratio of type-1 AGN, spectral analysis of the MgII emission lines was carried out to determine the black hole mass of type-1 AGN. For type-2 AGN, the Eddington ratio was determined using the blackhole mass which was inferred by assuming a $M_{\rm BH}-M_{\rm stellar}$ relationship where the stellar mass was estimated using SED fitting. The column density of the nuclear absorption of the AGN was estimated from the X-ray hardness ratio assuming a phenomenological X-ray spectral model.  The important results are summarised in the following bullets. We caution the reader that the black hole mass of type-2 AGN, as well as their Eddington ratio, were derived under the assumption that type-2 AGN follow the same black hole mass stellar mass as type-1 AGN.

\begin{enumerate}
    \item The Eddington ratio distribution of type-1 AGN is above the effective Eddington limit of dusty material whereas most type-2 AGN (75\%) are below the effective Eddington ratio limit. The observed Eddington ratio distribution of type-2 AGN is skewed towards a lower value than type-1. This implies that the line-of-sight obscuration of most type-2 AGN at redshift 0.8-1.8 is due to a radiation-pressure regulated dusty torus. (section \ref{sec:res:nhledd}).
    \item The X-ray absorption of type-1 AGN can be realised with mild ISM obscuration  \lognh{< 22} but for a part of the type-1 AGN sample with larger hydrogen column density, dust-free sight lines or obscuration by disk-winds are needed (section \ref{sec:disc:t1obs}).
    \item The fraction of type-2 AGN in the forbidden zone increased from $11\pm3$\% in the local universe to $28\pm5$\% at redshift 0.8-1.8. The increased fraction of AGN in the forbidden zone supports the increasing incidence of ISM absorption due to the cosmological evolution of galaxy ISMs which can be maintained under the AGN radiation pressure (section \ref{sec:disc:t2obs}).
    \item There is no trend between star-formation activity and obscuration. The total SFR of AGN host galaxies is consistent with typical main sequence galaxies in the same redshift or lower but not fully quenched (below 1\% SFR-MS). The SFR and SSFR of type-1 and type-2 AGN are consistent with each other at a 95\% confidence level. (section \ref{sec:res:sfa})
\end{enumerate}


\section*{Acknowledgements}

The authors would like to thank the anonymous reviewers for their helpful comments which greatly improved the quality of the manuscript. MA is supported by the Japan Society for the Promotion of Science (JSPS) KAKENHI Grant Number JP17H06129 and JP21H05583

The Hyper Suprime-Cam (HSC) collaboration includes the astronomical communities of Japan and Taiwan, and Princeton University.  The HSC instrumentation and software were developed by the National Astronomical Observatory of Japan (NAOJ), the Kavli Institute for the Physics and Mathematics of the Universe (Kavli IPMU), the University of Tokyo, the High Energy Accelerator Research Organization (KEK), the Academia Sinica Institute for Astronomy and Astrophysics in Taiwan (ASIAA), and Princeton University.  Funding was contributed by the FIRST program from the Japanese Cabinet Office, the Ministry of Education, Culture, Sports, Science and Technology (MEXT), the Japan Society for the Promotion of Science (JSPS), Japan Science and Technology Agency  (JST), the Toray Science  Foundation, NAOJ, Kavli IPMU, KEK, ASIAA, and Princeton University.

This paper is based [in part] on data collected at the Subaru Telescope and retrieved from the HSC data archive system, which is operated by Subaru Telescope and Astronomy Data Center (ADC) at NAOJ. Data analysis was in part carried out with the cooperation of Center for Computational Astrophysics (CfCA) at NAOJ.  We are honored and grateful for the opportunity of observing the Universe from Mauna kea, which has the cultural, historical and natural significance in Hawaii.

This paper makes use of software developed for Vera C. Rubin Observatory. We thank the Rubin Observatory for making their code available as free software at http://pipelines.lsst.io/. 

The Pan-STARRS1 Surveys (PS1) and the PS1 public science archive have been made possible through contributions by the Institute for Astronomy, the University of Hawaii, the Pan-STARRS Project Office, the Max Planck Society and its participating institutes, the Max Planck Institute for Astronomy, Heidelberg, and the Max Planck Institute for Extraterrestrial Physics, Garching, The Johns Hopkins University, Durham University, the University of Edinburgh, the Queen’s University Belfast, the Harvard-Smithsonian Center for Astrophysics, the Las Cumbres Observatory Global Telescope Network Incorporated, the National Central University of Taiwan, the Space Telescope Science Institute, the National Aeronautics and Space Administration under grant No. NNX08AR22G issued through the Planetary Science Division of the NASA Science Mission Directorate, the National Science Foundation grant No. AST-1238877, the University of Maryland, Eotvos Lorand University (ELTE), the Los Alamos National Laboratory, and the Gordon and Betty Moore Foundation.

These data were obtained and processed as part of the CFHT Large Area U-band Deep Survey (CLAUDS), which is a collaboration between astronomers from Canada, France, and China described in Sawicki et al. (2019, [MNRAS 489, 5202]).  CLAUDS is based on observations obtained with MegaPrime/ MegaCam, a joint project of CFHT and CEA/DAPNIA, at the CFHT which is operated by the National Research Council (NRC) of Canada, the Institut National des Science de l’Univers of the Centre National de la Recherche Scientifique (CNRS) of France, and the University of Hawaii. CLAUDS uses data obtained in part through the Telescope Access Program (TAP), which has been funded by the National Astronomical Observatories, Chinese Academy of Sciences, and the Special Fund for Astronomy from the Ministry of Finance of China. CLAUDS uses data products from TERAPIX and the Canadian Astronomy Data Centre (CADC) and was carried out using resources from Compute Canada and Canadian Advanced Network For Astrophysical Research (CANFAR).

Based on data obtained from the ESO Science Archive Facility with DOI(s): 	https://doi.org/10.18727/archive/58.

This research has made use of the NASA/IPAC Infrared Science Archive, which is funded by the National Aeronautics and Space Administration and operated by the California Institute of Technology.

\section*{Data Availability}

All catalogue data, Best-fitted SEDs, and future updates will be provided through \url{https://bovornpratch.github.io/} under the resource section.  Access to raw data will be provided upon reasonable request to the corresponding author.



\bibliographystyle{mnras}
\bibliography{ref} 




\appendix

\section{Catalog Data Desciption}
\label{sec:appendix}

Table \ref{tab:datatable} describes the columns in the supplement catalogue for 666 AGN used in this work. An example of the table is shown in table \ref{tab:exampletable}. For the optical photometric data, we refer to \citet{2022ApJ...941...97V}.

\begin{table*}
    \centering
    \caption{Description of supplement data catalogue}
    \label{tab:datatable}
    \begin{tabular}{ccc}
    \hline
    Name & unit & Description \\
    \hline
    XID\_XSERVS &  & X-ray source ID \\
    XCEN\_RA & deg & X-ray centroid right ascension (J2000) \\
    XCEN\_DEC & deg & X-ray centroid declination (J2000) \\
    OIR\_RA & deg & optical/ir counterpart right ascension (J2000)\\
    OIR\_DEC & deg & optical/ir counterpart declination (J2000) \\
    patch &  & HSC patch identifier \\
    tract &  & HSC tract identifier \\
    z &  & redshift \\
    z\_flag &  & redshift flag. 0=spec-z, 1=photo-z \\
    NIRCAT &  & VIDEO H-band detected flag \\
    IFRATIO &  &  HSC $i$-band flux ratio\\
    TYPE &  &  AGN classification \\
    swire\_irac3\_flux & uJy & SWIRE IRAC channel 3 flux density from datafusion catalogue \\
    swire\_irac3\_fluxunc & uJy & SWIRE IRAC channel 3 flux density uncertainty  \\
    swire\_irac4\_flux & uJy & SWIRE IRAC channel 4 flux density from datafusion catalogue \\
    swire\_irac4\_fluxunc & uJy & SWIRE IRAC channel 4 flux density uncertainty  \\
    swire\_mips24\_flux & uJy & SWIRE MIPS 24\um flux density from datafusion catalogue \\
    swire\_mips24\_fluxunc & uJy & SWIRE MIPS 24\um flux density uncertainty \\
    spire\_250\_flux & mJy &  SPIRE 250\um flux density from datafusion catalogue \\
    spire\_250\_fluxunc & mJy & SPIRE 250\um flux density uncertainty \\
    spire\_350\_flux & mJy & SPIRE 350\um flux density from datafusion catalogue \\
    spire\_350\_fluxunc & mJy & SPIRE 350\um flux density uncertainty \\
    spire\_500\_flux & mJy & SPIRE 500\um flux density from datafusion catalogue \\
    spire\_500\_fluxunc & mJy & SPIRE 500\um flux density uncertainty \\
    SB\_DETML &  & 0.5-2 keV band detection likelihood \\
    HB\_DETML &  & 2-10 keV band detection likelihood  \\
    logNH & [$\mathrm{cm^{-2}}$] & Hydrogen column density \\
    logNH\_LL & [$\mathrm{cm^{-2}}$] & Hydrogen column density lower uncertainty \\
    logNH\_UL & [$\mathrm{cm^{-2}}$] & Hydrogen column density upper uncertainty \\
    NH\_FLAG &  & Hydrogen column density flag (-99=upper/lower limit) \\
    logL\_210 & [$\mathrm{erg\ s^{-1}}$] & 2-10 keV X-ray luminosity \\
    logL\_210\_unc & [$\mathrm{erg\ s^{-1}}$] & 2-10 keV X-ray luminosity uncertainty \\
    logLBol & [$\mathrm{erg\ s^{-1}}$] & Bolometric luminosity \\
    logLBol\_unc & [$\mathrm{erg\ s^{-1}}$] & Bolometric luminosity uncertainty \\
    reduced\_chi\_square &   & Reduced $\chi^2$ from CIGALE \\
    logMbh & [$\mathrm{M_\odot}$] & Black hole mass \\
    logMbh\_lunc & [$\mathrm{M_\odot}$] & Black hole mass lower uncertainty \\
    logMbh\_uunc & [$\mathrm{M_\odot}$] & Black hole mass upper uncertainty  \\
    logMbh\_source &  & Source of black hole mass \\
    loglamledd &  & Eddington ratio \\
    loglamedd\_lunc &  & Eddington ratio lower uncertainty \\
    loglamedd\_uunc &  & Eddington ratio upper uncertainty \\
    logL6 & [$\mathrm{erg\ s^{-1}}$] & 6\um AGN luminosity \\
    logL6\_unc & [$\mathrm{erg\ s^{-1}}$] & 6\um AGN luminosity uncertainty \\
    logMs & [$\mathrm{M_\odot}$] & Host galaxy stellar mass \\
    logMs\_unc & [$\mathrm{M_\odot}$] & Host galaxy stellar mass uncertainty \\
    logsfr & $[\mathrm{M_\odot\ yr^{-1}}]$ & Star-formation rate \\
    logsfr\_unc & $[\mathrm{M_\odot\ yr^{-1}}]$ & Star-formation rate uncertainty \\
    \hline
\end{tabular}
\end{table*}

\begin{table*}
\centering
\caption{Example of table \ref{tab:datatable}. The full table is available online.}
\label{tab:exampletable}
\begin{tabular}{cccccccccc}
\hline
XID\_XSERVS & OIR\_RA & OIR\_DEC & TYPE & logNH & logL\_210 & logMbh & logLBol & logMs & logsfr \\
\hline
 & Deg & Deg &  & $[\mathrm{cm^{-2}}]$ & $[\mathrm{erg s^{-1}}]$ & $[\mathrm{M_{\odot}}]$ & $[\mathrm{erg s^{-1}}]$ & $[\mathrm{M_{\odot}}]$ & $[\mathrm{M_{\odot} yr^{-1}}]$ \\
 \hline
XMM00003 & 34.201198 & -4.498737 & 2 & 22.57 & 43.63 & 8.48 & 44.78 & 10.82 & 1.95 \\
XMM00013 & 34.208313 & -5.295092 & 1 & 22.04 & 43.66 & 8.00 & 44.86 & 10.51 & 1.66 \\
XMM00023 & 34.213887 & -5.418938 & 2 & 22.64 & 43.52 & 8.11 & 44.65 & 10.50 & 0.13 \\
XMM00034 & 34.220582 & -5.183233 & 1 & 20.00 & 44.16 & 8.60 & 45.50 & 10.27 & -1.78 \\
XMM00039 & 34.223993 & -5.348011 & 2 & 22.63 & 43.48 & 7.98 & 44.61 & 10.39 & 0.81 \\
\hline
\end{tabular}
\end{table*}


\bsp	
\label{lastpage}
\end{document}